\pdfoutput=1
\documentclass[final,5p,times,twocolumn]{elsarticle}

\usepackage{graphicx}
\usepackage{caption}
\usepackage{subcaption}
\usepackage{soul}
\usepackage{pgfplots}
\usepackage{epstopdf}
\usepackage{multirow}
\usepackage{listings}
\usepackage{fancybox}
\usepackage{color}
\usepackage{colortbl}
\usepackage{algorithm}
\usepackage{algorithmic}
\usepackage{tcolorbox}% http://ctan.org/pkg/tcolorbox
\makeatletter
\newcommand{\mybox}[1]{%
  \setbox0=\hbox{#1}%
  \setlength{\@tempdima}{\dimexpr\wd0+13pt}%
  \begin{tcolorbox}[boxrule=0.5pt, colback=white, arc=4pt,
	        left=6pt,right=6pt,top=6pt,bottom=6pt,boxsep=0pt]
			    #1
  \end{tcolorbox}
}

\definecolor{mygray}{gray}{.9}

\definecolor{mycolorhigh}{RGB}{168, 157, 89}
\definecolor{mycolormiddle}{RGB}{224, 224, 182}
\definecolor{mycolorlow}{RGB}{245, 245, 237}

\newcommand{\tool}{SETU}
\newcommand{\projects}{12}
\newcommand{\reports}{3,689}
\newcommand{\company}{Baidu CrowdTest}

\journal{Information and Software Technology}

\begin{document}

\begin{frontmatter}


\title{Cutting Away the Confusion From Crowdtesting}



\author{Junjie Wang$^{1,3}$, Mingyang Li$^{1,3}$, Song Wang$^4$, Tim Menzies$^5$, Qing Wang$^{1,2,3,*}$}

\address{$^1$Laboratory for Internet Software Technologies, $^2$State Key Laboratory of Computer Science, \\
Institute of Software Chinese Academy of Sciences, Beijing, China\\
$^3$University of Chinese Academy of Sciences, Beijing, China, $^*$Corresponding author \\
$^4$Electrical and Computer Engineering, University of Waterloo, Canada\\
$^5$Department of Computer Science, North Carolina State University, Raleigh, NC, USA\\
Email: \{wangjunjie,limingyang,wq\}@itechs.iscas.ac.cn, song.wang@uwaterloo.ca, tim@menzies.us}

\begin{abstract}
%should use structured abstract

\textbf{Context}: Crowdtesting is effective especially when it comes to the feedback on GUI systems, or subjective opinions about features.
Despite of this, we find crowdtesting reports are highly duplicated, i.e., 82\% of them are duplicates of others.
Most of the existing approaches mainly adopted textual information for duplicate detection, and suffered from low accuracy because of the lexical gap.
Our observation on real industrial crowdtesting data found that when dealing with crowdtesting reports of GUI systems, the reports would be accompanied with images, i.e., the screenshots of the tested app. 
We assume the screenshot to be valuable for duplicate crowdtesting report detection because it reflects the real context of the bug and is not affected by the variety of natural languages.

\textbf{Objective}: We aim at automatically detecting duplicate crowdtesting reports that could help reduce triaging effort.

\textbf{Method}: In this work, we propose {\tool} which combines information from the ScrEenshots and the TextUal descriptions to detect duplicate crowdtesting reports.
We extract four types of features to characterize the screenshots (i.e., image structure feature and image color feature) and the textual descriptions (i.e., TF-IDF feature and word embedding feature), and design a hierarchical algorithm to detect duplicates based on the four similarity scores derived from the four features respectively.

\textbf{Results}: We investigate the effectiveness of {\tool} on {\projects} projects with {\reports} reports from one of the Chinese largest crowdtesting platforms.
Results show that recall@1 achieved by {\tool} is 0.44 to 0.79, recall@5 is 0.66 to 0.92, and MAP is 0.21 to 0.58 across all experimental projects.
Furthermore, {\tool} can outperform existing state-of-the-art approaches significantly and substantially.

\textbf{Conclusion}: Through combining the screenshots and textual descriptions, our proposed {\tool} can improve the duplicate crowdtesting reports detection performance. 

\end{abstract}


\end{frontmatter}


% \input{sec/1-introduction.tex}
% \input{sec/2-background.tex}
% \input{sec/3-relatedWork.tex}
% \input{sec/4-approach.tex}
% \input{sec/5-design.tex}
% \input{sec/6-result.tex}
% \input{sec/7-discussion.tex}
% \input{sec/8-conclusion.tex}

\section{Introduction}
\label{sec:intro}

Crowdtesting is an emerging trend in software testing which accelerates testing processes by attracting online crowd workers to accomplish various types of testing tasks \cite{crowdsourced_ESEM2016,crowd_test_report_prioritization,junjie_ase2016,crowd_report_prioritization_ase2016,junjie_icse2017,gao2018successes}.
It entrusts testing tasks to crowd workers whose diverse testing environments/platforms, background, and skill sets could significantly contribute to more reliable, cost-effective, and efficient testing results.

The benefit of crowdtesting must be carefully assessed with respect to the cost of the technique. 
At first place, crowdtesting is a scalable testing method under which large software systems can be tested with appropriate results.
This is particular true when the testing is related with the feedback on GUI systems, or subjective opinions about different features. 

One aspect of crowdtesting which is not received enough attention in prior work is the confusion factors in crowdtesting results.
Our observation on real industrial data shows that an average of
82\% crowdtesting reports are duplicate, which suggests much of the crowdtesting work can be optimized.
A significant problem with such a large number of duplicate reports is that the subsequent analysis by software testers becomes extremely complicated. 
For example, we find that merely working through 500 crowdtesting reports to find the duplicate ones takes almost the whole working day of a tester.  
This paper mostly removes that effort by a novel method for detection of duplicate reports.

The issue of duplicate reports has been studied in terms of textual descriptions~\cite{runeson2007detection,wang2008approach,jalbert2008automated,sun2010discriminative,sureka2010detecting,sun2011towards,prifti2011detecting,tian2012improved,banerjee2012automated,zhou2012learning,duplicate_topic_model,banerjee2013fusion,alipour2013contextual,hindle2016contextual,duplicate_SANER_empirical,duplicate_embedding} (see details in Section \ref{sec:related}). 
However, in practice, it is common that different people might use different terminologies, or write about different phenomena to describe the same issue \cite{wang2008approach,duplicate_topic_model,tian2012improved,duplicate_embedding}, which makes
the descriptions often confusing. Because of this, most existing approaches for duplicate report detection suffer from low accuracy.
However, when dealing with crowdtesting reports of GUI systems, besides the textual descriptions, often the feedback is in the form of images. 
Our observation on real industrial crowdtesting data reveals that an average of 94\% crowdtesting reports are accompanied with an image, i.e., screenshot of the app.
We suppose this is another valuable source of information for detecting duplicate crowdtesting reports.
Compared with the textual description, a screenshot can reflect the real context of the bug and is not affected by the variety of natural languages.

In this paper, we propose {\tool} which combines information from the ScrEenshots and the TextUal descriptions to detect duplicate crowdtesting reports.
We first extract two types of features from screenshots (i.e., image structure feature and image color feature), and two types of features from textual descriptions (i.e., TF-IDF feature and word embedding feature).
We then obtain the screenshot similarity and textual similarity through computing the similarity scores based on the four types of features. 
To decide the duplicates of a query report, {\tool} adopts a hierarchical algorithm.
% whether two given reports are duplicates with each other, 
Specifically, if the screenshot similarity between the query report and candidate report is higher than a threshold, we treat the candidate report as the first class and rank all reports in the first class by their textual similarity.
Otherwise, we treat it as the second class (follow behind the first class) and rank all reports in this class by their combined textual similarity and screenshot similarity.
Finally, we return a list of candidate duplicate reports of the query report, with the ranked reports of the first class followed by the ranked reports of the second class.

We experimentally evaluate the effectiveness of {\tool} on {\projects} projects with {\reports} crowdtesting reports from one of the Chinese largest crowdtesting platforms.
Results show that the recall@1 achieved by {\tool} is 0.44 to 0.79, recall@5 is 0.66 to 0.92, and MAP is 0.21 to 0.58 across all experimental projects.
These results significantly and substantially outperform three state-of-the-art and typical duplicate detection approaches.
In addition, we also experimentally evaluate the necessity of screenshots and textual descriptions in duplicate detection, as well as the relative effect of the four types of features.

This paper makes the following contributions:
\begin{itemize}
\item We show that the screenshots are valuable in duplicate crowdtesting reports detection and we further demonstrate the need to use both the screenshots and the textual descriptions in detecting duplicate crowdtesting reports.

\item We propose a novel approach ({\tool}) for duplicate crowdtesting report detection, which combines the information from screenshots and textual descriptions hierarchically.

\item We evaluate the effectiveness of {\tool} on {\projects} projects from one of the Chinese largest crowdtesting platforms, and results are promising.

\end{itemize}

The rest of this paper is organized as follows:
Section \ref{sec:background} describes the background and motivation of this study, while Section \ref{sec:related} surveys related work.
Section \ref{sec:approach} presents our proposed approach of duplicate detection.
Section \ref{sec:experiment} and \ref{sec:result} show the experimental setup and evaluation results respectively.
Section \ref{sec:discussion} provides a detailed discussion and threats to validity.
Finally, we summarize this paper in Section \ref{sec:conclusion}.

% Before beginning, we digress one definition. 
% Prior work on exploring duplicated bug reports use the term \textit{duplication}.
% Here we will say \textit{replication}, and we will reserve the term duplication for simple exploration of text-based documents.
% Instead, we will explore methods for finding duplicated documents with screenshots. 

\section{Background and Motivation}
\label{sec:background}

\subsection{Background}
\label{subsec:background_crowdsourced}

In this section, we present a brief background of crowdtesting to help better understand the challenges we meet in real industrial crowdtesting practice.

\begin{figure}[ht!]
\centering
\includegraphics[width=7cm]{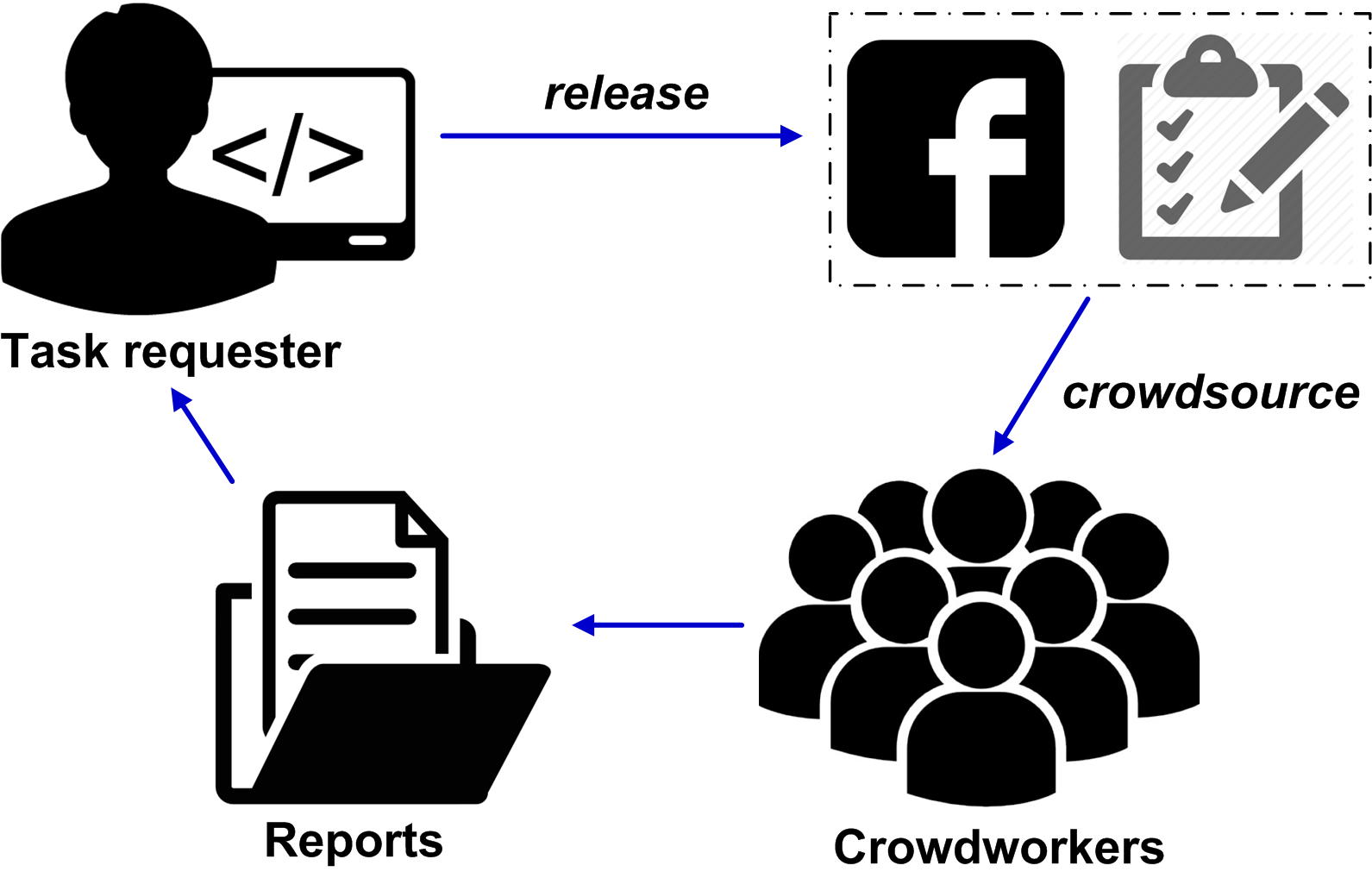}
\caption{Procedure of crowdtesting}
\label{fig:crowdtesting}
\vspace{-0.1in}
\end{figure}

Our experiment is conducted with {\company} crowdtesting platform\footnote{Baidu (baidu.com) is the largest Chinese search service provider. Its crowdtesting platform (test.baidu.com) is also one of the largest crowdtesting platforms in China.}.
As shown in Figure \ref{fig:crowdtesting}, in general, the task requester\footnote{The task requesters are usually the testers in practice, so we also call them as the testers in the following paper.} prepares the software under test and testing tasks, and distributes them on the crowdtesting platform.
Then, the crowd workers can sign in to conduct the tasks and are required to submit the crowdtesting reports, i.e., to describe the process and result of the testing task s/he carried on, which include the input, operation steps, result description, and screenshot.
Table \ref{tab:report_example} demonstrates an example of the crowdtesting report\footnote{Note that, since all the reports are written in Chinese in our experiment projects, we translate them into English to facilitate understanding.}.
% The platform can automatically record a crowd worker's information
% and the environment on
% which a testing task is carried on.

In order to attract more workers, testing tasks are often financially compensated. Under this context, workers can submit hundreds of reports for a crowdtesting task.
This platform delivers approximately 100 projects per month,
and receives more than 1,000 test reports per day on average.
Among these reports, an average of 82\% crowdtesting reports are
duplicates of other reports.

Currently in this platform, the testers need to manually inspect these crowdtesting reports to identify the duplicate ones.
However, inspecting 500 reports manually could take almost the whole
working day of a tester.
Obviously, such process is time-consuming and low-efficient.

\begin{table}[!t]
\scriptsize
\caption{An example of crowdtesting report}
\label{tab:report_example}
\centering\scalebox{0.85}{
\begin{tabular}{p{1.8cm}|p{6cm}}
\hline
\textbf{Attribute} &  \textbf{Description: \textit{example}}  \\
\hline
\textbf{Environment} & Phone type: \textit{Samsung SN9009}
              Operating system: \textit{Android 4.4.2}
              ROM information: \textit{KOT49H.N9009}
              Network environment: \textit{WIFI} \\
\hline
\textbf{Crowd worker} & Id: \textit{123456}
              Location: \textit{Beijing Haidian District} \\
\hline
\textbf{Testing task} & Id: \textit{01}
                  Name: \textit{Incognito mode} \\
\hline
\textbf{Input and operation steps} & \textit{Input ``sina.com.cn'' in the browser, then click the first news. Select ``Setting'' and then set ``Incognito Mode''. Click the second news in the website. Select ``Setting'' and then select ``History''. }\\
\hline
\textbf{Result description} & \textit{``Incognito Mode'' does not work as expected. The first news, which should be recorded, does not appear in ``History''.} \\
\hline
\textbf{Screenshot} & \includegraphics[width=1in]{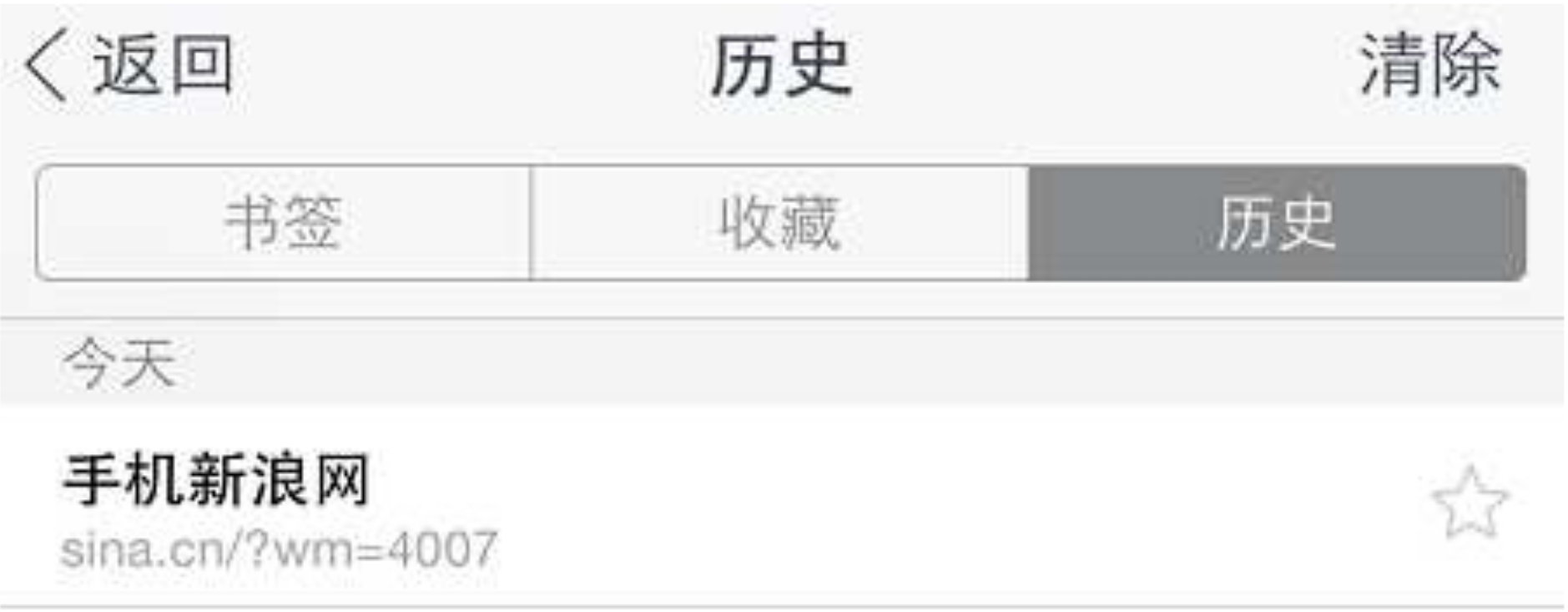} \\
\hline
\textbf{Assessment} & Passed or failed given by crowd worker: \textit{Failed} \\
\hline
%\textbf{Judgement} & Whether it is a true fault judged by the tester in company: \textit{True fault} \\
%\hline
\end{tabular}
}
\vspace{-0.05in}
\end{table}

\subsection{Motivation}
\label{subsec:background_motivation}

In this section, we present two examples from {\company} crowdtesting platform to motivate the need of using
both the screenshots and the textual descriptions in duplicate crowdtesting report detection.
These examples draw from a sport application,
i.e., JIAJIA Sport.
% \footnote{http://www.anzhi.com/pkg/f024_com.edergen.android.ijumpapp.html}.
It can automatically record and analyze users' sport-related information such as running, rope skipping, etc.
To better test this application, its testers distribute the testing task on {\company} crowdtesting platform, and 462 crowdtesting reports are submitted by the crowd workers.
By analyzing these crowdtesting reports, we find the following two motivating examples.

\subsubsection{\textbf{Motivating Example 1: Descriptions Could be Confusing}}

Crowdtesting reports \textit{Rope-145} and \textit{Rope-270} are about the sharing function.
Their descriptions are as follows:

\textbf{\textit{Rope-145: I press the qzone\footnote{qzone is a popular social networking website in China.} sharing button in the bottom and want to share my rope skipping record, but nothing happens.}}

\textbf{\textit{Rope-270: I press the sharing button and want to share my rope skipping record to qzone, but nothing happens except a failure notice.}}

\begin{figure}[!ht]
  \centering
  \hspace{-0.1in}
  \begin{subfigure}{0.24\textwidth}
    \includegraphics[width=4.5cm]{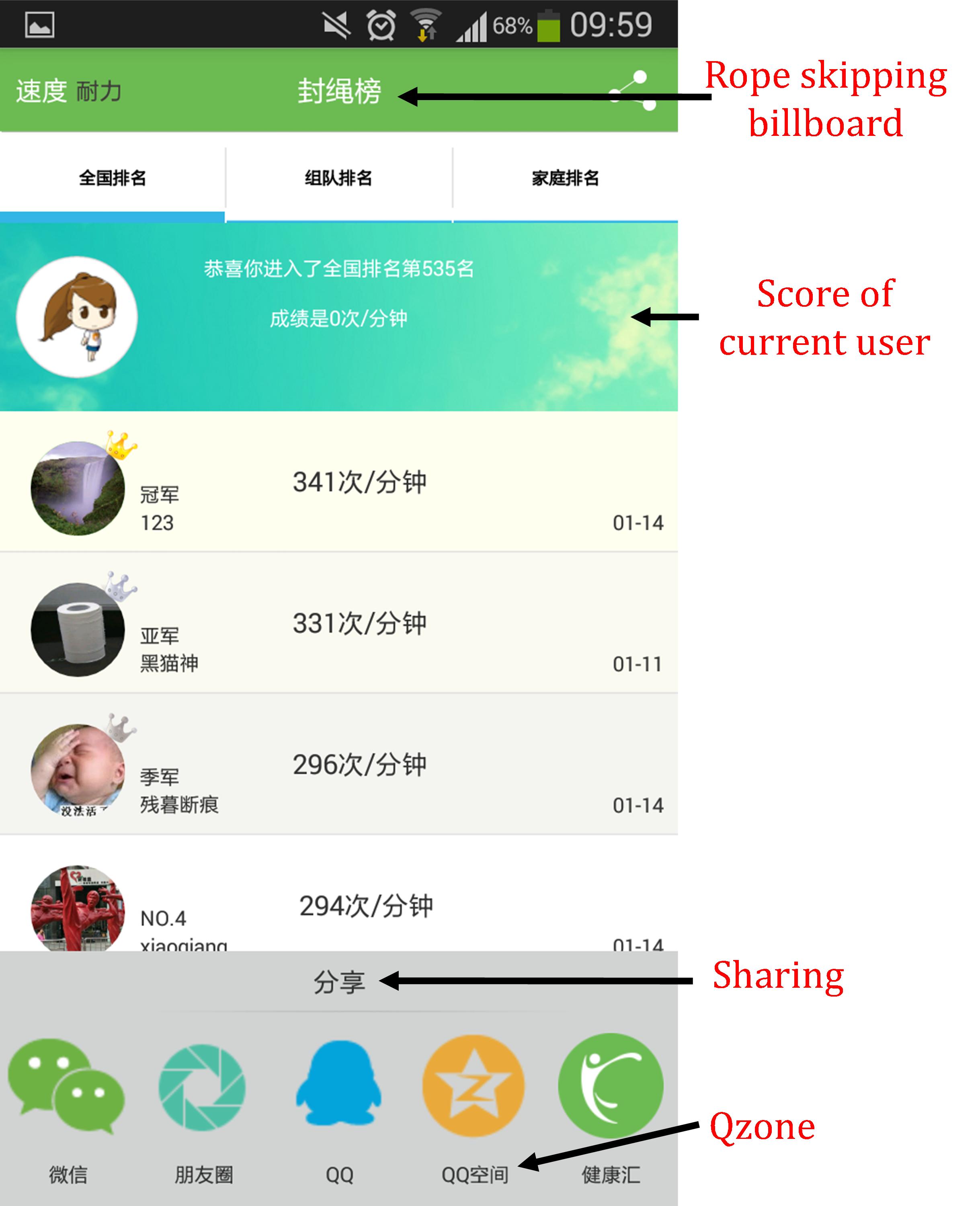}
	 \caption{Rope-145}
	 \label{fig:M1-1}
  \end{subfigure}
  \hspace{-0.05in}
  \begin{subfigure}{0.24\textwidth}
    \includegraphics[width=4.5cm]{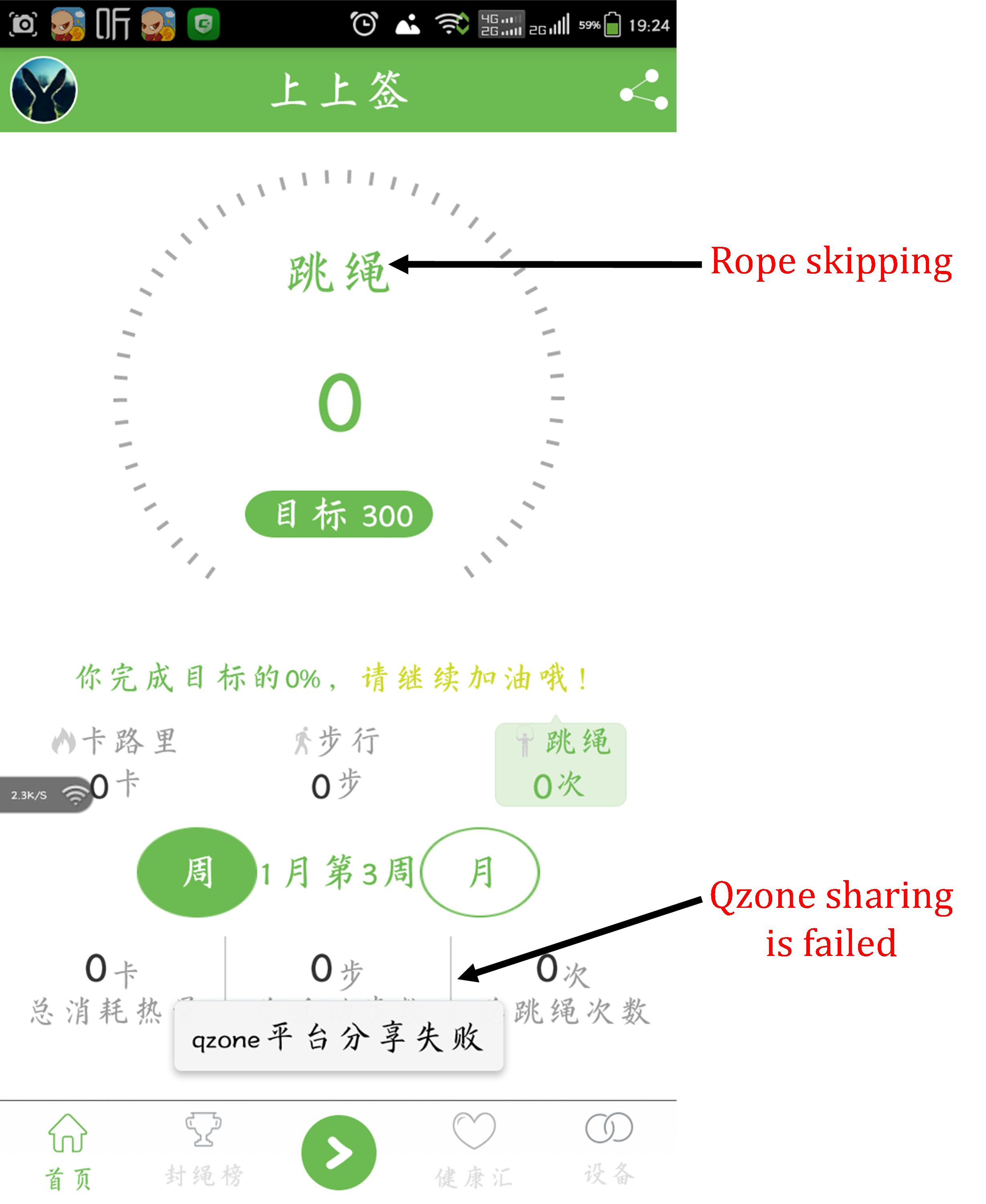}
	\caption{Rope-270}
  \label{fig:M1-2}
  \end{subfigure}
  \caption{Motivating example 1}
  \label{fig:motivation_1}
\end{figure}

Both descriptions contain such words as ``sharing button'', ``rope skipping record'', and ``qzone''.
Using traditional duplicate detection approaches,
these two reports would be identified as duplicates with a high probability.
However, if the screenshot information (see Figure \ref{fig:motivation_1}) is considered, they can be easily determined as non-duplicates, which is the ground truth.
The screenshot of \textit{Rope-145} is about the billboard of rope skipping record, and the crowdtesting report reveals a bug about sharing the ranking of rope skipping record.
For the screenshot of \textit{Rope-270}, it demonstrates the detail page of rope skipping record, and the report reveals a bug about sharing the detailed record.
In this sense, the screenshot provides the context-related information and can help better detect duplicate reports.
One may argue that the above information should be in the \textit{operation steps} (see Table \ref{tab:report_example}) submitted by the crowd workers.
However, the crowd workers are far from professional testers, and in our datasets only few reports contain the detailed and correct operation steps.
% \yang{The two sentences above look abrupt here. The fact that ``only 30\% reports contain the detailed operation steps'' seems more likely introducing the background. Is it better to put the fact in section background and add a sentence ``Like many reports in platform, Rope-145 and Rope-270 don't give operations steps, and only descriptions given by testers are as follows:'' at the start of Motivation Example 1?}

\mybox{\textbf{Finding 1}: The textual descriptions of crowdtesting reports may easily lead to confusing understanding.
With the help of context-related information provided by screenshots, the duplicate crowdtesting reports
can be detected more accurately.}

\begin{table*}[!t]
\scriptsize
\caption{\textbf{Summary of duplicate bug report detection researches}}
\label{tab:relatedWork}
\centering{
\begin{tabular}{p{4cm}|p{5.8cm}|p{1.2cm}|p{2.5cm}|p{0.4cm}|p{1.2cm}}
\hline
\textbf{Article} & \textbf{Basic idea} & \textbf{Experiment projects} & \textbf{Performance} & \textbf{Baselines} & \textbf{Publication year} \\
\hline
\cite{runeson2007detection} Detection of Duplicate Defect Reports Using Natural Language Processing & Use natural language processing technique (i.e., tokenization, stemming, stop words removal, vector space representation) to compute the similarity & Sony Ericsson Mobile communications & Recal@5 = 30\%,   Recall@10 = 38\%,   Recall@15 = 42\% & N/A & ICSE'2007 \\
\hline
\cite{wang2008approach} An Approach to Detecting Duplicate Bug Reports using Natural Language and Execution Information  & Use both natural language information and execution information to compute the similarity, and combine the two similarity values based on heuristics & Firefox & Recall@1 = 67\%,   Recall@10 = 93\% & \cite{runeson2007detection} & ICSE'2008  \\
\hline
\cite{jalbert2008automated} Automated Duplicate Detection for Bug Tracking Systems  & Build a machine learning classifier which combines
the surface features of the report, textual similarity metrics, and graph clustering algorithms & Mozilla & Recall@1 = 25\%,   Recall@5 = 38\%,   Recall@10 = 45\% & \cite{runeson2007detection} & DSN'2008 \\
\hline
\cite{sun2010discriminative} A Discriminative Model Approach for Accurate Duplicate Bug Report Retrieval & Extract 54 features to measure the textual similarity of a pair of reports based on the term weighting, then build a machine learning classifier on these features & Firefox, Eclipse, OpenOffice & Recall@1 = 32\% - 38\%,   Recall@5 = 48\% - 52\%,   Recall@10 = 56\% - 61\% & \cite{runeson2007detection}, \cite{wang2008approach}, \cite{jalbert2008automated} &  ICSE'2010 \\ 
\hline
\cite{sureka2010detecting} Detecting Duplicate Bug Report Using Character N-Gram-Based Features  & Compute the semantic and lexical similarity based on  the character-level n-gram model
& Eclipse & Recall@10 = 40\%,   Recall@20 = 48\%,   Recall@50 = 61\% & N/A & APSEC'2010 \\
\hline
\cite{sun2011towards} Towards More Accurate Retrieval of Duplicate Bug Reports  & Use both the textual information and other fields information (e.g., product, component, versions) to measure the similarity, extend BM25F (an effective similarity formula) to handle lengthy structured report by considering weights of terms & Eclipse, OpenOffice, Mozilla & Recall@1 = 37\% - 42\%,   Recall@5 = 58\% - 62\%,   Recall@10 = 63\% - 69\%,   MAP = 45\% - 53\% & \cite{sun2010discriminative}, \cite{sureka2010detecting} & ASE'2011 \\
\hline
\cite{prifti2011detecting} Detecting Bug Duplicate Reports through Local Reference & Find the reports whose submit time is within the sliding time window, then rank these reports based on TF-IDF & Firefox & Recall@1 = 20\%,   Recall@5 = 36\%,   Recall@10 = 44\% & N/A & PROMISE'2011 \\
\hline
\cite{tian2012improved} Improved Duplicate Bug Report Identification  & Use extended BM25F \cite{sun2011towards} for similarity measurement, introduce relative similarity by considering the top-k most similar reports (rather than only considering top-1 most similar report) & Mozilla & Recal@4 = 24\%,   Recall@20 = 25\% & \cite{jalbert2008automated} & CSMR'2012 \\
\hline
\cite{banerjee2012automated} Automated Duplicate Bug Report Classification using Subsequence Matching  & 
Hypothesize that two reports that have the longest ordered sequence of common words are more likely to be duplicates than those have the same frequency of unordered words; Propose common sequence matching approach & Firefox & Recall@1 = 30\%,    Recall@5 = 51\%,    Recall@10 = 58\% & N/A & HASE'2012 \\
\hline 
\cite{zhou2012learning} Learning to Rank Duplicate Bug Reports & Identify 9 textual and statistical features of bug reports, create a ranking model based on these features to rank the duplicate bugs higher, learn the weights of the features by applying the stochastic gradient descent algorithm & Eclipse & Recall@1 = 49\%,    Recall@5 = 69\%,    Recall@10 = 76\%,   MRR = 58\% & \cite{sun2010discriminative},  \cite{sun2011towards} & CIKM'2012 \\
\hline
\cite{duplicate_topic_model} Duplicate Bug Report Detection with a Combination of
Information Retrieval and Topic Modeling  & Use information retrieval and topic modeling techniques to measure the textual similarity & Eclipse, OpenOffice, Mozilla & Recall@1 = 42\% - 57\%,   Recall@5 = 66\% - 76\%,   recall@10 = 80\% - 86\% & \cite{sun2011towards} & ASE'2012 \\
\hline
\cite{banerjee2013fusion} A Fusion Approach For Classifying Duplicate
Problem Reports  & Use multi-label classification to assign each report with multiple duplicate prediction results and use a fusion role to combine the results & Firefox & recall@1 = 31\%,   recall@5 = 50\%,   recall@10 = 58\% & \cite{sun2010discriminative} & ISSRE'2013 \\
\hline
\cite{alipour2013contextual}\cite{hindle2016contextual} A Contextual Approach Towards More Accurate Duplicate
Bug Report Detection and Ranking  & Utilize contextual information about software-quality attributes, software-architecture terms, and system-development topics when computing the similarity & Eclipse, OpenOffice, Mozilla, Android & MAP = 29\% - 51\% & \cite{sun2011towards} & MSR'2013, EMSE'2016 \\
\hline
\cite{duplicate_SANER_empirical} An Empirical Study on Recommendations of Similar Bugs & Use bug component field and textual similarity of bug descriptions to measure the similarity & Mozilla & Recall@1 = 36\%,   Recall@3 = 43\% &  \cite{sun2011towards} & SANER'2016 \\
\hline 
\cite{duplicate_embedding} Combining Word Embedding with Information Retrieval to Recommend Similar Bug Reports & Apply information retrieval and word embedding technique to measure the textual similarity of bug titles and descriptions, and consider bug product and component fields in the final similarity of two reports & Eclipse, Mozilla & Recall@1 = 16\% - 19\%,   Recall@5 = 37\% - 43\%,   Recall@10 = 48\% - 53\%,   MAP = 26\% - 33\%,   MRR = 48\% - 59\% & \cite{duplicate_SANER_empirical} & ISSRE'2016 \\
\hline
\end{tabular}
}
  
\scriptsize{Note that, \cite{hindle2016contextual} is the extension of \cite{alipour2013contextual}, so we put them together and present the results of the extended paper \cite{hindle2016contextual}}.
\end{table*}

\subsubsection{\textbf{Motivating Example 2: Screenshots Usually Lack Details}}

In this example, crowdtesting reports~\textit{Rope-62} and \textit{Rope-217}
have similar screenshots (see Figure \ref{fig:motivation_2}).
If the duplicate detection is only based on the screenshot information,
these two reports would be determined as duplicates with a high probability.
However, the ground truth is just the opposite.
Their descriptions are as follows:

\begin{figure}[!ht]
  \centering
  \hspace{-0.1in}
  \begin{subfigure}{0.24\textwidth}
    \includegraphics[width=4.5cm]{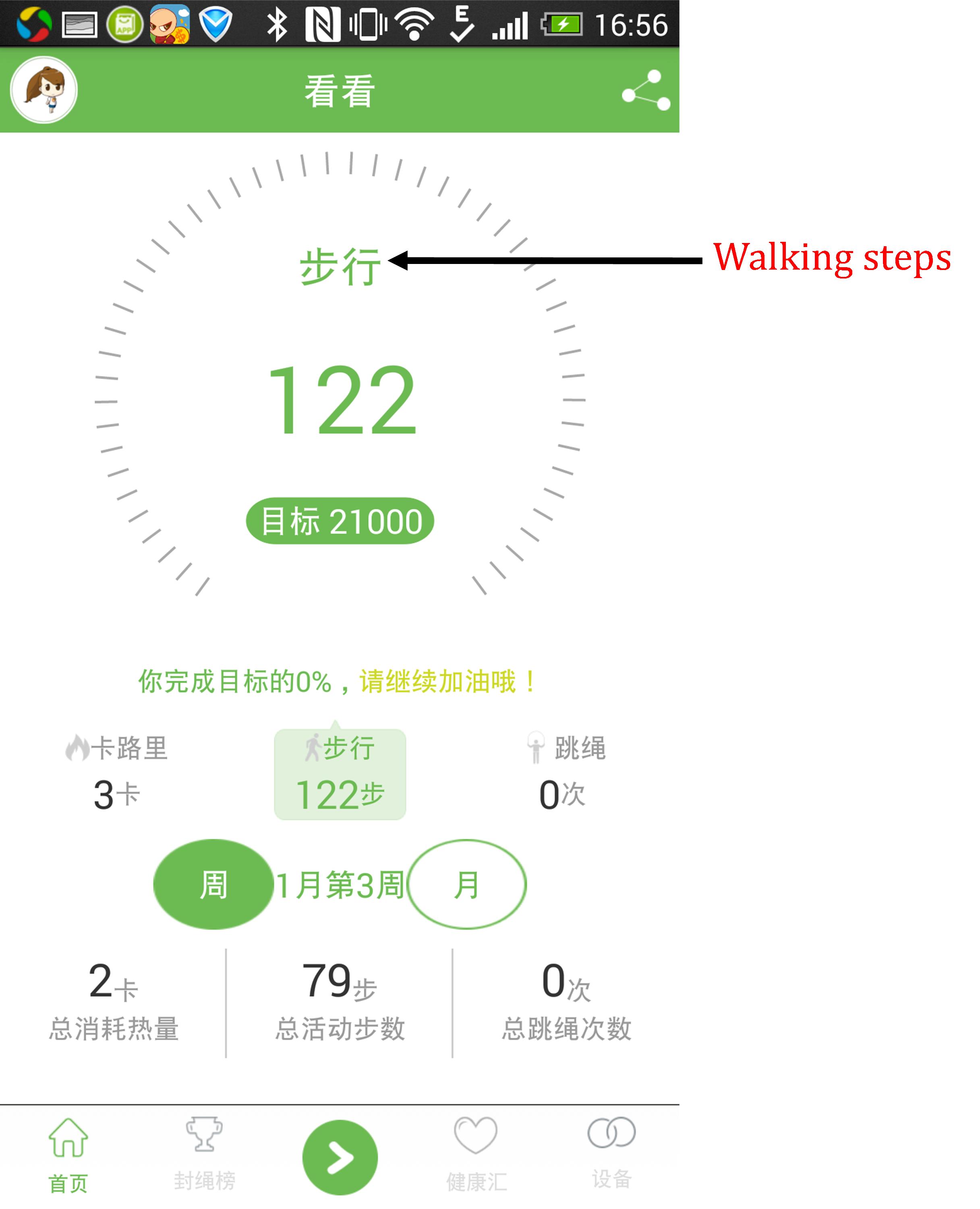}
	 \caption{Rope-62}
	 \label{fig:M2-1}
  \end{subfigure}
  \hspace{-0.05in}
  \begin{subfigure}{0.24\textwidth}
    \includegraphics[width=4.5cm]{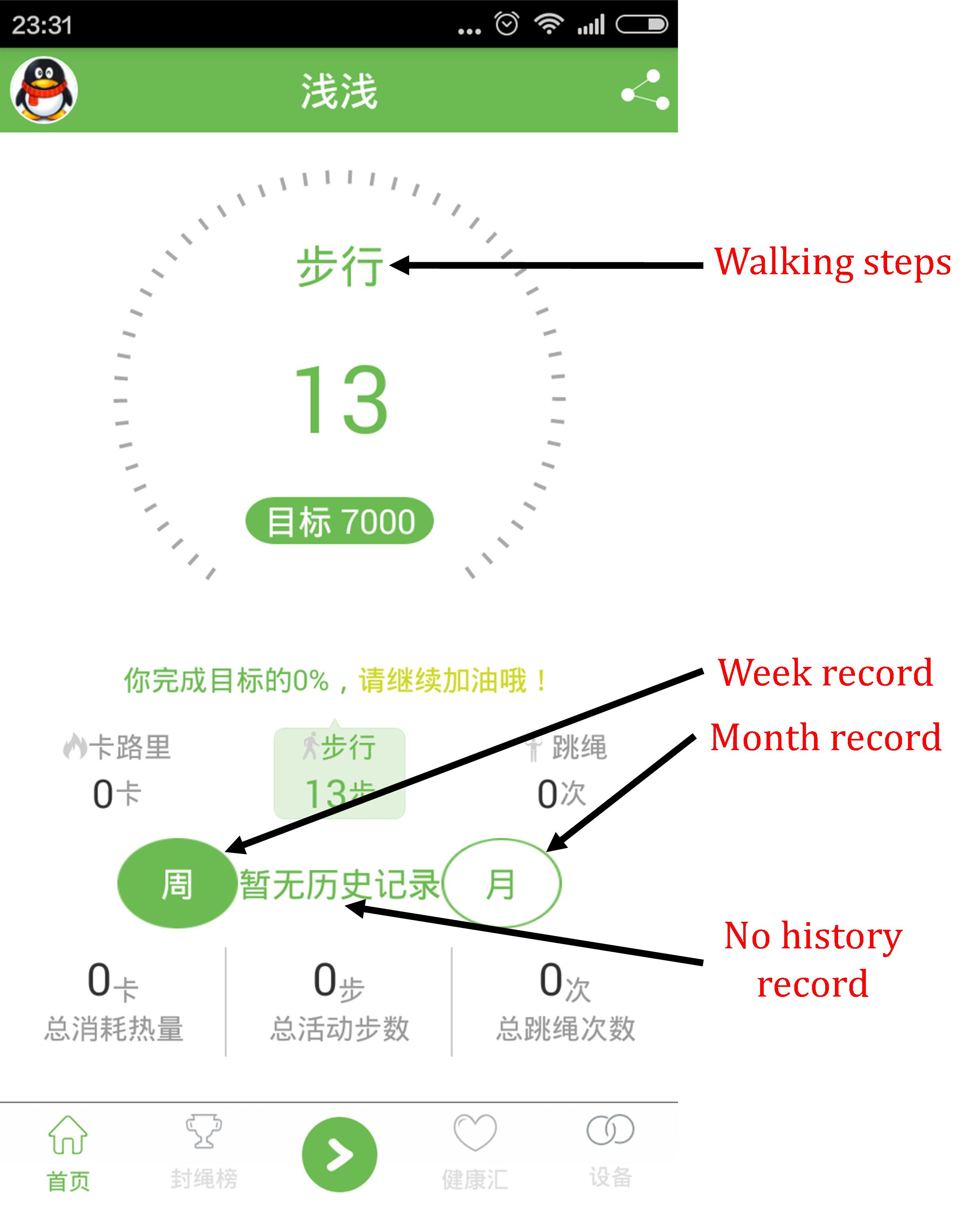}
	\caption{Rope-217}
  \label{fig:M2-2}
  \end{subfigure}
  \caption{Motivating example 2}
  \label{fig:motivation_2}
\end{figure}

\textbf{\textit{Rope-62: I walked for 10 minutes, but the steps only increased by 10. An hour later, i just sat on my chair, but the steps increased sharply. }}

\textbf{\textit{Rope-217: In the detail page, there is indeed step record for today, but there is no step record for this week.}}

From the descriptions, we can easily observe that these two crowdtesting reports involve two different bugs, although under the same function which is denoted by the two highly similar screenshots.
In this sense, a screenshot merely demonstrates the context-related information about a crowdtesting report.
We still need to refer to the textual descriptions to finally determine whether they are duplicates.

\mybox{\textbf{Finding 2}: Screenshots of crowdtesting reports mainly demonstrate the context-based information. 
Under a specific context, only with the detailed illustration provided by the textual description, the duplicated reports can be accurately detected.}

% \yang{Example 1 uses a negative example and shows that screen shot can help distinguish if two reports refer to the same bug. Therefore, shoots can be a filter.    Example 2 shows, although shoots give contexts, we still need textual descriptions to finally determined. In example 2, I feel shoots are easily to be similar, that is also reason why we use words to finally determined. Example 2 can motivate us to use textual descriptions in first class.  But I think it may be necessary to use another example to explain why use both two similarities in second class. }

% \yang{If possible, I think it is better to use another positive example that two  reports are truly duplicates and have lexical gap, the results are more convincing with the help of shoots. It can motive us to use average of tow similarities in our algorithm.}

\section{Related Work}
\label{sec:related}

\subsection{Crowdtesting}
\label{subsec:related_crwodsourced}

Crowdtesting has been applied to facilitate many testing tasks.
% e.g., test case generation~\cite{crowd_test_puzzle_test_case_generation,dolstra2013crowdsourcing,mao2017crowd,rojas2017code}, usability testing~\cite{crowd_test_usability_test}, software performance analysis~\cite{crowd_test_performance}, software bug detection and reproduction~\cite{mobile16}.
Chen and Kim~\cite{crowd_test_puzzle_test_case_generation} applied
crowdtesting to test case generation.
They investigated object mutation and constraint solving issues underlying existing
test generation tools, and presented a puzzle-based automatic testing environment.
Musson et al. \cite{crowd_test_performance} proposed an approach, in which the crowdworker
was used to measure real-world performance of software products.
Gomide et al. \cite{crowd_test_usability_test} proposed an approach that employed
a deterministic automata to help usability testing.
Adams et al.~\cite{mobile16} proposed MoTIF to detect and reproduce crashes in mobile apps after their deployment in the wild.

These studies leverage crowdtesting to solve the problems in traditional testing activities,
while some other approaches focus on the new encountered problem in crowdtesting.

Feng et al. \cite{crowd_test_report_prioritization,crowd_report_prioritization_ase2016}
proposed approaches to prioritize test reports in crowdtesting.
They designed strategies to dynamically select the most risky and diversified test
report for inspection in each iteration.
Jiang et al. \cite{jiang2018fuzzy} proposed the test report fuzzy clustering framework by aggregating redundant and multi-bug crowdtesting reports into clusters to reduce the number of inspected test reports.
Wang et al.~\cite{crowdsourced_ESEM2016,junjie_ase2016,junjie_icse2017} proposed approaches to automatically classify crowdtesting reports. Their approaches can overcome
the different data distribution among different software domains,
and attain good classification results.
Liu et al. \cite{liu2018generating} proposed an automatic approach to generate descriptive words for the screenshots based on the language model and Spatial Pyramid Matching technique. 
Cui et al. \cite{cuiqiang_seke2017,cuiqiang_compsac2017} and Xie et
al.~\cite{xie2017cocoon} proposed crowd worker selection approaches to recommend
appropriate crowd workers for specific crowdtesting tasks. These
approaches considered the worker's experience, relevance
with the task, diversity of testing context, etc., and recommend a
set of workers who can detect more bugs.

In this work, we focus on detecting the duplicated crowdtesting reports to facilitate
real industrial crowdtesting practice.

\begin{figure*}[ht!]
\centering
\includegraphics[width=17cm]{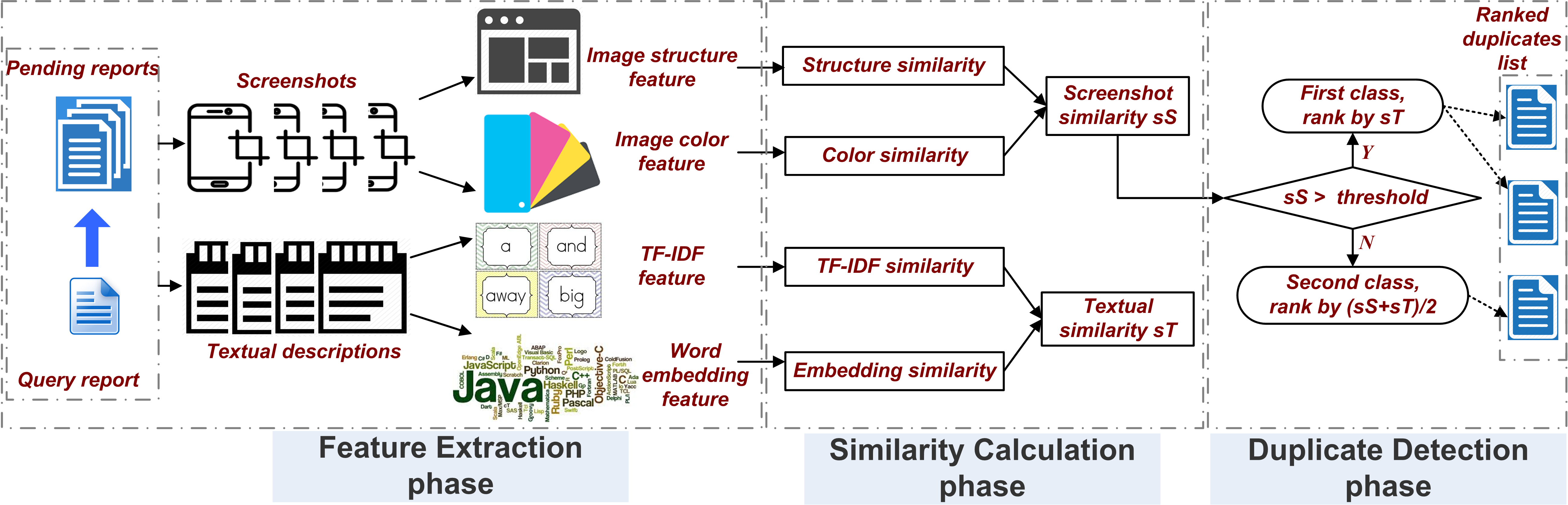}
\caption{Overview of {\tool}}
\label{fig:overview}
\end{figure*}

\subsection{Duplicated Bug Report Detection}
\label{subsec:related_dupIssue}

Many approaches have been proposed to detect duplicate bug reports~\cite{runeson2007detection,wang2008approach,jalbert2008automated,sun2010discriminative,sureka2010detecting,sun2011towards,prifti2011detecting,tian2012improved,banerjee2012automated,zhou2012learning,duplicate_topic_model,banerjee2013fusion,alipour2013contextual,hindle2016contextual,duplicate_SANER_empirical,duplicate_embedding}.
The main focus of duplicate report detection is to obtain the similarity of two reports. 
Runeson et al.~\cite{runeson2007detection} proposed the first duplicate report detection approach which uses natural language information of bug reports to compute the similarity. 
Later, other sources of information were utilized in similarity measurement, e.g., execution trace \cite{wang2008approach}, fields information as product and component \cite{tian2012improved,duplicate_SANER_empirical,duplicate_embedding}.
Meanwhile, techniques to improve the similarity measurement precision were also introduced, e.g., BM25F similarity measurement \cite{sun2011towards}, topic modeling \cite{duplicate_topic_model}, word embedding technique \cite{duplicate_embedding}, etc.

Table \ref{tab:relatedWork} presents a summary of existing duplicate bug report detection researches.

% \song{Improving the accuracy of duplicate bug report detection using textual similarity measures MSR'14}

% \song{Mining duplicate questions in stack overflow MSR'16}

These approaches mainly leverage the textual information and fields information to detect the duplicate reports.
In this work, we introduce the screenshot information to facilitate the duplicate crowdtesting report detection.
Our proposed approach combines the information from both the screenshots and the textual descriptions, and can detect duplicate crowdtesting reports with high accuracy.

To evaluate the effectiveness of our proposed approach (in Section \ref{sec:experiment} and \ref{sec:result}), we choose three state-of-the-art and typical approaches \cite{duplicate_topic_model,duplicate_SANER_empirical,duplicate_embedding} as the baselines. 
The reason why we use \cite{duplicate_SANER_empirical,duplicate_embedding} is because they are the latest two researches which can be treated as the state-of-the-art approaches. 
The reason why we use \cite{duplicate_topic_model} is because they can achieve the highest performance across all existing researches. 
We had planned to include \cite{hindle2016contextual}, which is also one of the latest researches, as baseline. 
However, the contextual words utilized in that work are English-oriented; simply translating these words to Chinese or translating our crowdtesting reports to English would bring serious information loss. 
Therefore, this paper utilize three state-of-the-art and typical approaches \cite{duplicate_topic_model,duplicate_SANER_empirical,duplicate_embedding} as baseline in evaluation (see details in Section \ref{subsec:experiment_baseline}).

\section{Approach}
\label{sec:approach}

Motivated by the two examples in Section \ref{subsec:background_motivation}, we propose a duplicate detection approach ({\tool}), 
which combines information from both the ScrEenshots and the TextUal descriptions to detect duplicate crowdtesting reports.
Figure \ref{fig:overview} illustrates the overview of {\tool}.

Our approach is organized as a pipeline comprising three phases: feature extraction, similarity calculation, and duplicate detection.

In the feature extraction phase, for the screenshots, we extract two types of features, i.e., image structure feature and image color feature (details are in Section~\ref{subsec:approach_image}).
For the textual descriptions, we also extract two types of features, i.e., TF-IDF (Term Frequency and Inverse Document Frequency) feature and word embedding feature (details are in Section~\ref{subsec:approach_text}).

In the similarity calculation phase, based on the four types of features,
we compute four similarity scores between the query report and each of the pending reports,
and obtain the screenshot similarity and textual similarity.
Cosine similarity, which is commonly used for measuring distance \cite{crowdsourced_ESEM2016,runeson2007detection,duplicate_SANER_empirical,data_ming}, is employed in our approach.
% We will not introduce this phase in the following subsections as it shares the same procedure as the above studies.

In the duplicate detection phase (details are in Section~\ref{subsec:approach_detection}), we design a hierarchical algorithm. 
In detail, if the screenshot similarity between the query report and pending report is larger than a specific threshold, we treat the pending report as first class and rank all reports in the first class by their textual similarity.
Otherwise, we treat it as second class (follow behind the first class) and rank all reports in this class by their combined textual similarity and screenshot similarity.
Finally, we return a list of candidate duplicate reports of the query report, with the ranked reports of the first class followed by the ranked reports of the second class. 
The reason why we separate reports in two classes is to
take the advantages of the information provided by screenshots.
As described in Section \ref{subsec:background_motivation}, the screenshots can provide the context-related information, and reports with different screenshots are very unlikely to be duplicate with each other. We will further discuss the performance of other combinations of the screenshot similarity and textual similarity in Section \ref{subsec:dis_alternative}.

%As we described in Section~\ref{sec:intro},
%compared with textual description, screenshot information can
%reflect the real scenario of the issues and is not affected by the
%variety of natural languages.
%The assumption behind our tool is that duplicated test reports have a high
%probability to share similar screenshots.
%take a note about whether they share similar screenshots,
%which will be used in case study (Section \ref{subsec:result_rq3}).
%Moreover, we also need to ensure all reports in first class are ranked higher than second class.

\subsection{Extracting Screenshot Features}
\label{subsec:approach_image}

We extract screenshot features from the image of screenshot accompanied with each crowdtesting report.
Note that, in our experimental projects, all the crowdtesting reports contain zero or one screenshot (with details in Section \ref{subsec:experiment_data}). 
For reports which do not have a screenshot, we use a default blank picture to extract the features.
Future work will consider the situation that one report contains several screenshots. 
We use the following two types of features.

\subsubsection{\textbf{Image Structure Feature}}

The geometric structure of an image exhibits fundamental information for distinguishing screenshots \cite{image_role_lifeifei,image_role}.
Images with highly similar geometric structures (e.g., line segments) would be probably the same screenshot.

Gist descriptor \cite{image_structure_feature} can capture the spatial structure of an image through the segmentation and processing of individual regions.
We use the publicly available package\footnote{\small{http://people.csail.mit.edu/torralba/code/spatialenvelope/}} with default parameters to extract image structure feature.
It results in a 128-dimensional image structure feature vector for each image.

\begin{figure}[!ht]
  \centering
  \hspace{-0.1in}
  \begin{subfigure}{0.24\textwidth}
    \includegraphics[width=4.5cm]{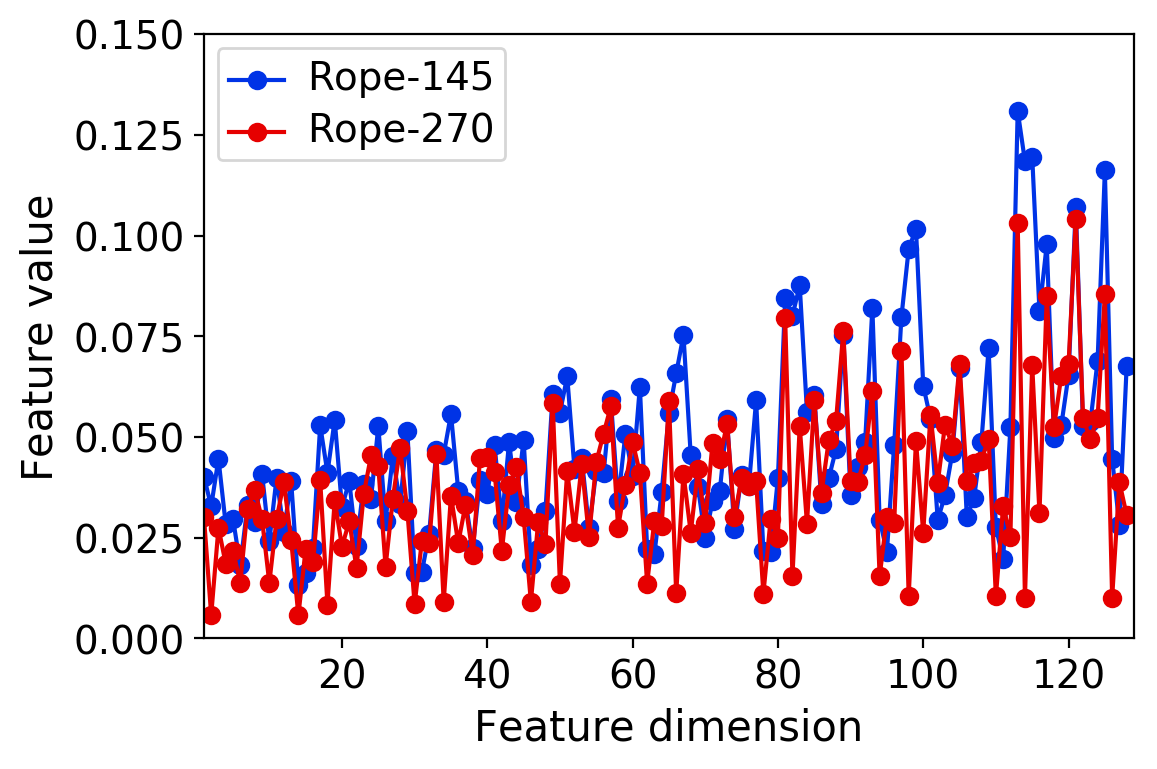}
	 \caption{Rope-145 and Rope-270}
	 \label{fig:structure-1}
  \end{subfigure}
  \hspace{-0.05in}
  \begin{subfigure}{0.24\textwidth}
    \includegraphics[width=4.5cm]{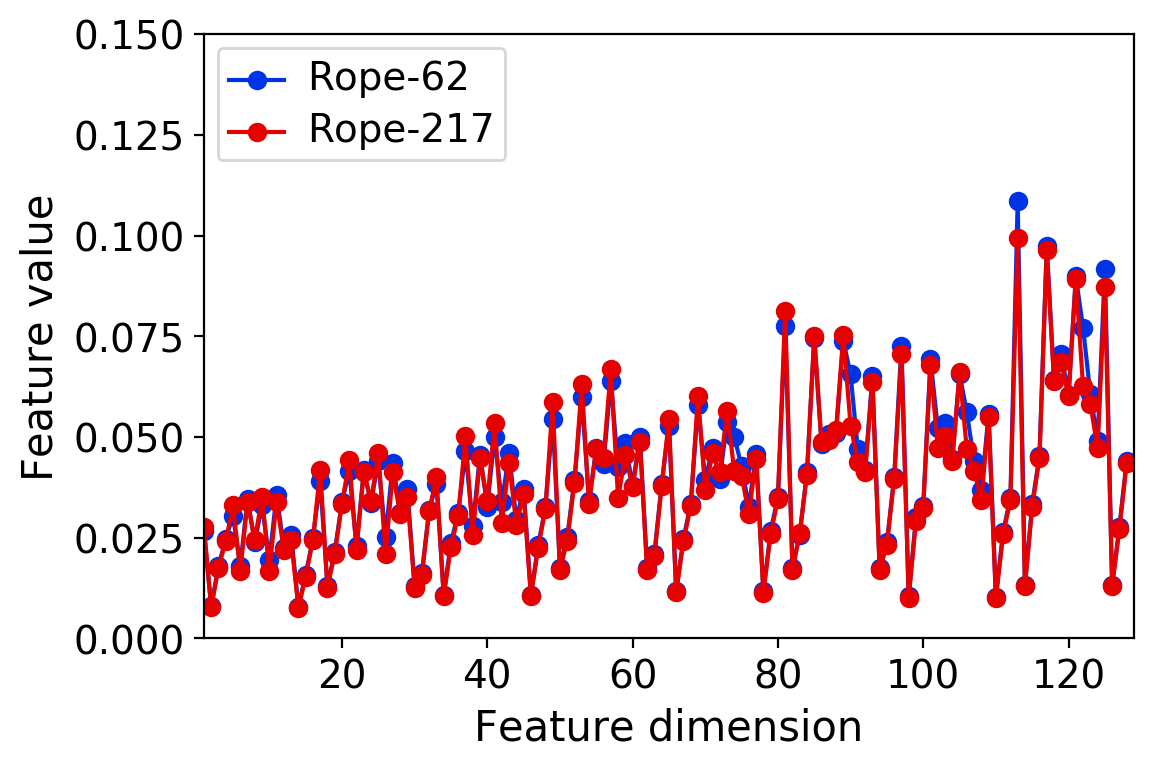}
	\caption{Rope-62 and Rope-217}
  \label{fig:structure-2}
  \end{subfigure}
  \caption{Image structure features}
  \label{fig:structure}
\end{figure}

Figure \ref{fig:structure} shows the image structure features for the four screenshots in Figure \ref{fig:motivation_1} and Figure \ref{fig:motivation_2}.
The $x$ axis is the feature's dimension (i.e., 128),
while the $y$ axis is the value of each dimension.
We can easily find that the image structure feature vectors of \textit{Rope-145} and \textit{Rope-270} are obviously different with each other, while the structure feature vectors of \textit{Rope-62} and \textit{Rope-217} are almost the same.
This coincides with our visual perception.

\subsubsection{\textbf{Image Color Feature}}

Color is another basic indicator of visual contents, and is often used to describe and represent an image \cite{image_role_lifeifei,image_role}.
Images with highly similar color distributions are probably the same screenshot.

MPEG-7 descriptor \cite{image_color_feature} can capture the representative colors on a grid superimposed of an image.
Specifically it is designed to extract the spatial distribution of color in an image through grid based representative color selection and transformation.
We also use the publicly available software\footnote{\small{http://www.semanticmetadata.net/}} to extract the image color feature.
It results in a 189-dimensional image color feature vector for each image.

\begin{figure}[!ht]
  \centering
  \hspace{-0.1in}
  \begin{subfigure}{0.24\textwidth}
    \includegraphics[width=4.5cm]{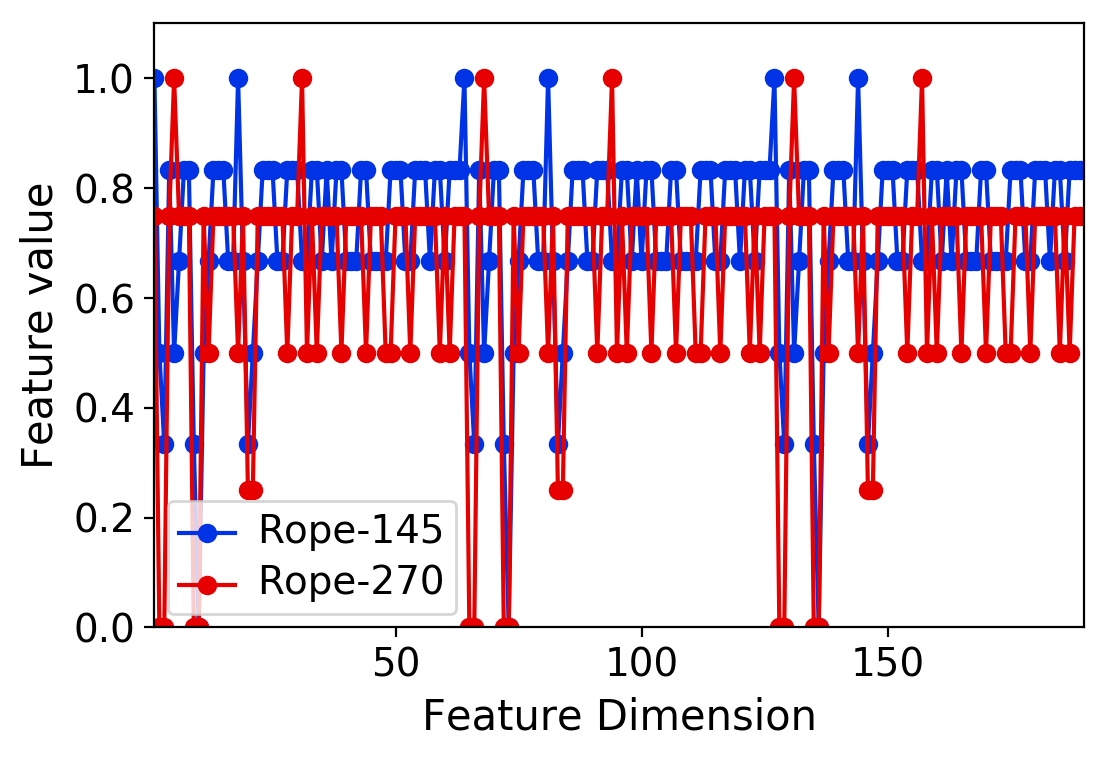}
	 \caption{Rope-145 and Rope-270}
	 \label{fig:color-1}
  \end{subfigure}
  \hspace{-0.05in}
  \begin{subfigure}{0.24\textwidth}
    \includegraphics[width=4.5cm]{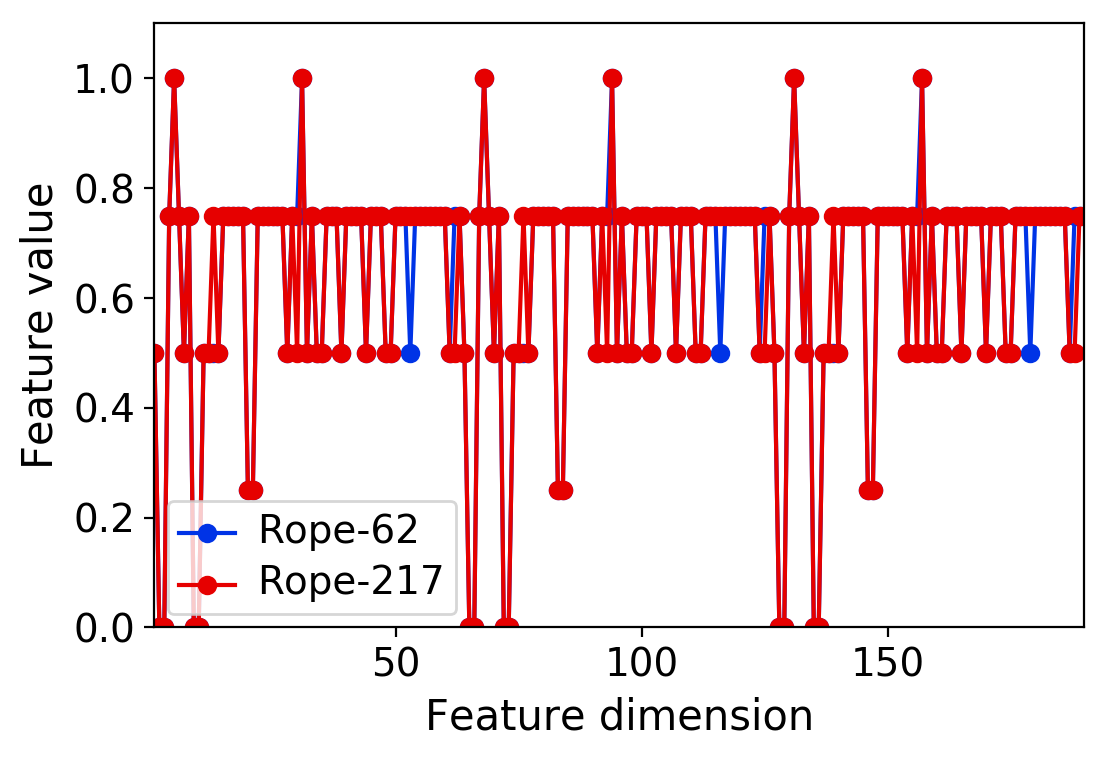}
	\caption{Rope-62 and Rope-217}
  \label{fig:color-2}
  \end{subfigure}
  \caption{Image color features}
  \label{fig:color}
\end{figure}

Figure \ref{fig:color} demonstrates the image color feature vectors
for the four screenshots in Figure \ref{fig:motivation_1} and Figure \ref{fig:motivation_2}.
The image color feature vector of \textit{Rope-145} exerts obvious difference with the vector of \textit{Rope-270}, while the color vector of \textit{Rope-62} is quite similar with the vector of \textit{Rope-217}.
Just as the image structure feature, this coincides with our visual perception.

Both structure feature and color feature are widely used in image processing tasks~\cite{image_role_lifeifei,image_role}, so we adopt both of them in the duplicate crowdtesting report detection and explore their performance in Section~\ref{sec:result}.

\subsection{Extracting Textual Features}
\label{subsec:approach_text}

We extract textual features from the textual descriptions of crowdtesting reports.

We first collect different sources of textual descriptions together
(input, operation steps, and result description),
and then conduct the natural language processing to remove noise and extract core terms.
Specifically, because the crowdtesting reports in our experiment are written in Chinese,
we adopt ICTCLAS\footnote{ICTCLAS (http://ictclas.nlpir.org/) is widely used Chinese NLP platform.} for word segmentation, and segment descriptions into words.
We then remove \textbf{stopwords} (i.e., ``am'', ``on'', ``the'', etc.) to reduce noise.
Note that, crowd workers often use different words to express the same concept, so we introduce the \textbf{synonym replacement} technique to mitigate this problem.
Synonym library of LTP\footnote{LTP (http://www.ltp-cloud.com/) is considered as one of the best cloud-based Chinese NLP platforms.} is adopted.
The remaining terms are saved and will be used to extract the following two types of features.

\subsubsection{\textbf{TF-IDF Feature}}
\label{subsubsec:approach_tfidf}

TF-IDF (Term Frequency and Inverse Document Frequency) is one of the most popular feature for representing textual documents in information retrieval.
The main idea of TF-IDF is that if a term appears many times in one report and a few times in the other report, the term has a good capability to differentiate the reports, and thus the term has high TF-IDF value.
Specifically, given a term $t$ and a report $r$, $TF(t,r)$ is the number of times that term $t$ occurs in report $r$, while $IDF(t)$ is obtained by dividing the total number of reports by the number of reports containing term $t$.
TF-IDF is computed as: $TF-IDF(t,r) = TF(t,r) \times IDF(t)$.

%\vspace{-0.1in}
% \begin{equation}
% \label{equation_tf}
% TF(t,r) = \frac{\mathit{number \ of \ times \ t \ appears \ in \ r}}{\mathit{number \ of \ terms \ in \ r}}
% \end{equation}
%
%\vspace{-0.1in}
%\begin{equation}
%\label{equation_idf}
%IDF(t) = \log \frac{total \ number \ of \ reports}{number \ of \ reports \ that \ contain \ t + 1}
%\end{equation}

%Finally, TF-IDF is computed as:
%
%\vspace{-0.1in}
%\begin{equation}
%\label{equation_tfidf}
%TF-IDF(t,r) = TF(t,r) \times IDF(t)
%\end{equation}

With the above formula, the textual description of a report $r$ can be represented as a TF-IDF vector, i.e., $r = (w_1, w_2, ..., w_n)$, where $w_i$ denotes the TF-IDF value of the $i^{th}$ terms in report $r$.

\subsubsection{\textbf{Word Embedding Feature}}

Word embedding is a feature learning technique in natural language processing where individual words are no longer treated as unique symbols, but represented as $d$-dimensional vector of real numbers that capture their contextual semantic meanings~\cite{embedding_model,embedding_model_Bengio}.

We use the publicly available software\footnote{https://code.google.com/archive/p/word2vec/} to obtain the word embedding of a report.
With the trained word embedding model, each word can be transformed into a $d$-dimensional vector where $d$ is set to 100 as suggested in previous studies~\cite{duplicate_embedding,duplicate_linkable_knowledge_stackoverflow}.
Meanwhile a crowdtesting report can be transformed into a matrix in which each row represents a term in the report.
We then transform the report matrix into a vector by averaging all the word vectors the report contains as previous work did \cite{duplicate_embedding}.
Specifically, given a report matrix that has $n$ rows in total, we denote the $i^{th}$ row of the matrix as $r_{i}$ and the transformed report vector $v_{d}$ is generated as follows:

\vspace{-0.1in}
\begin{equation}
\label{equation_3}
v_{d} = \frac{\sum_{i}r_{i}}{n}
\end{equation}

With the above formula, each crowdtesting report can be represented as a word embedding vector.

The TF-IDF feature focuses on the similarity of reports considering the term matching, while the word embedding feature concerns more on the relationship of terms considering the context they appear.
We adopt both of them in our approach and investigate their performance in duplicate report detection in Section~\ref{sec:result}.

\subsection{Conducting Duplicate Report Detection}
\label{subsec:approach_detection}

Following the previous studies~\cite{runeson2007detection,wang2008approach,jalbert2008automated,sun2010discriminative,sureka2010detecting,sun2011towards,prifti2011detecting,tian2012improved,banerjee2012automated,zhou2012learning,duplicate_topic_model,banerjee2013fusion,alipour2013contextual,hindle2016contextual,duplicate_SANER_empirical,duplicate_embedding}, our duplicate crowdtesting report detection problem is formulated as follows:
Given a query report of a crowdtesting project, our approach would recommend a list of duplicate reports from all the pending reports of the project and rank them by their probabilities to be duplicates.
We design a hierarchical algorithm (Algorithm~\ref{algorithm_approach}) to detect the duplicate reports.

\begin{algorithm}[htb]
\footnotesize
\caption{Duplicate report detection algorithm}
\label{algorithm_approach}
\hrule
\begin{algorithmic}[1]
\vspace{.2cm}
\REQUIRE ~\\
    Pending crowdtesting report set $R$; \\
    Query report $q$;
    Threshold $thres$
\ENSURE ~\\
    A list of duplicate reports $D$;
\FOR{each report $r$ in $R$ and $q$}
    \STATE Extract the image structure feature vector and image color feature vector from its screenshot;
    \STATE Extract the TF-IDF feature vector and word embedding feature vector from its textual description;
\ENDFOR
\FOR{each report $r$ in $R$}
    \FOR{$fType$ in [$structure$, $color$, $tfidf$, $embedding$]}
        \STATE Compute the cosine similarity between the $fType$ vector of $q$ and $r$, and denote as $s\_fType$;
    \ENDFOR
    %\STATE Compute the cosine similarity between the image structure vectors of $q$ and $r$ and denote as $s\_structure$;
%    \STATE Compute the cosine similarity between the image color vectors of $q$ and $r$ and denote as $s\_color$;
%    \STATE Compute the cosine similarity between the TF-IDF vectors of $q$ and $r$ and denote as $s\_tfidf$;
%    \STATE Compute the cosine similarity between the word embedding vectors of $q$ and $r$ and denote as $s\_embedding$;

    \STATE $s\_{\mathit{screenshot}} = (\mathit{s\_structure + s\_color)/2}$;
    \STATE $\mathit{s\_textual = (s\_tdidf + s\_embedding)/2}$;
    \STATE $\mathit{s\_total = (s\_screenshot + s\_textual)/2}$
    \IF{$\mathit{s\_screenshot} > \mathit{thres}$}       \STATE put $r$ in $\mathit{D\_first}$
    \ELSE        \STATE put $r$ in $\mathit{D\_second}$	
    \ENDIF
\ENDFOR
\STATE Rank $\mathit{D\_first}$ based on reports' $\mathit{s\_textual}$ and put them in $D$ sequentially
\STATE Rank $\mathit{D\_second}$ based on reports' $\mathit{s\_total}$ and put them in $D$ sequentially (follow behind the reports of $\mathit{D\_first}$)
\end{algorithmic}
\hrule
\end{algorithm}
%\vspace{-0.1in}

In the algorithm, the screenshot similarity can be seen as a filter because the screenshot provides the context-related information (see Section \ref{subsec:background_motivation} for details).
If two crowdtesting reports have different screenshots,
they are unlikely to be duplicate even if the textual similarity between them is high.
In addition, if two crowdtesting reports are accompanied with the same screenshot, whether they are duplicate reports mainly depends on the similarity of their textual descriptions.

The threshold to determine whether two screenshots are the same one is an input parameter. We explore the influence of this parameter on the detection performance in Section \ref{subsec:experiment_parameter}.

We have also experimented with different weights for $\mathit{s\_structure}$ and $\mathit{s\_color}$ when combining them to obtain $\mathit{s\_screenshot}$ (Line 9 in Algorithm \ref{algorithm_approach}), the weights for $\mathit{s\_tfidf}$ and $\mathit{s\_embedding}$ when combining them to get $\mathit{s\_textual}$ (Line 10 in Algorithm \ref{algorithm_approach}), as well as the weights for $\mathit{s\_screenshot}$ and $\mathit{s\_textual}$ when combining them to obtain $\mathit{s\_total}$ (Line 11 in Algorithm \ref{algorithm_approach}).
Results turned out that when the two similarities have an equal weight, {\tool} can achieve a relative good and stable performance. 
Due to space limit, we do not present the detailed results.
Note that, this does not imply the screenshot and textual descriptions are equally important, because the weights for  $\mathit{s\_screenshot}$ and $\mathit{s\_textual}$ are only used in the second class of reports (Line 10 and 19 in Algorithm \ref{algorithm_approach}). 
Another note is that, for the reports in the second class, we have experimented with other ranking manners, i.e., by $\mathit{s\_screenshot}$, by $\mathit{s\_textual}$, and results turned out that with the ranking manner shown above, the detection performance is relative good and stable. 
% This is reasonable since the reports in the second class are less likely to be the duplicates, different ranking manner might not bring too much difference in the performance. 

\section{Experiment Design}
\label{sec:experiment}

\subsection{Research Questions}
\label{subsec:experiment_rq}

Our evaluation addresses the following research questions:

\begin{itemize}
\item \textbf{RQ1 Effectiveness}: How effective is {\tool} in detecting duplicate crowdtesting reports?
\end{itemize}

RQ1 aims at evaluating the effectiveness of {\tool} in duplicate reports detection. 
We also compare {\tool} with the state-of-the-art approaches (see details in Section \ref{subsec:experiment_baseline}) to investigate whether and to what extent it improves over prior work. 

\begin{itemize}
\item \textbf{RQ2 Necessity}: Are both screenshots and textual descriptions necessary in detecting duplicate crowdtesting reports?
\end{itemize}

This paper proposes to utilize both screenshots and textual descriptions in duplicate detection. RQ2 is to investigate whether both of them are necessary. 
We employ two additional experiments to investigate it. 
See details in Section \ref{subsec:experiment_setup}.
% We employ two additional experiments, i.e., \textit{onlyText} and \textit{onlyImage}, to investigate this research question. 
% In detail, \textit{onlyText} denotes ranking the reports only based on the textual similarity (i.e., TF-IDF  similarity + word embedding similarity), while \textit{onlyImage} is based on the screenshot similarity (i.e., color similarity + structure similarity).

\begin{itemize}
\item \textbf{RQ3 Replaceability}: What is the relative effect of the four types of features (i.e., TF-IDF, word embedding, image color, and image structure) in detecting duplicate crowdtesting reports?
\end{itemize}

{\tool} employs four types of features to characterize the screenshots (i.e., image color and structure) and the textual descriptions (i.e., TF-IDF and word embedding). 
RQ3 is to investigate the relative effect of these features.
We use another four experiments for investigating this RQ, with details in Section \ref{subsec:experiment_setup}.
% We use another four experiments, i.e., \textit{noTF}, \textit{noEmb}, \textit{noClr} and \textit{noStrc}, to investigate this research question. 
% For example, \textit{noTF} denotes applying other three features except TF-IDF (i.e., only use embedding, image color and image structure feature) to conduct the duplicate detection. 

% \begin{itemize}
% \item \textbf{RQ4 Alternative}: Can other combination manner of screenshot similarity and textual similarity outperform {\tool} in detecting duplicate crowdtesting reports?
% \end{itemize}

% The motivating examples in Section \ref{subsec:background_motivation} indicate that
% the role of screenshots is mainly to demonstrate the context-related information, while the role of textual descriptions is to provide detailed illustration of the reported problem.  
% This is why our proposed approach {\tool} first uses screenshot similarity to filter the reports to first class and conducts the ranking separately. 

% However, one may still argue that other combination manners could achieve better performance.  
% RQ4 aims at investigating whether other combination manners could outperform {\tool} in duplicate detection. 
% We design three new combination manners to investigate this RQ. Details are shown in Section \ref{subsec:experiment_setup}.

\subsection{Experimental Dataset}
\label{subsec:experiment_data}

We mentioned that our experiment is based on crowdtesting reports from the repositories of {\company} crowdtesting platform.
We collect all crowdtesting projects closed between June 1st 2017 and June 10th 2017.
There are totally {\projects} crowdtesting projects. 

Table \ref{tab:projects} presents the detailed information of the projects with the application domain, the number of reports (i.e., \textit{\# report}), the number and percentage of reports which have screenshots (i.e., \textit{Num scr.} and \textit{Pert. scr.}).

There is a label accompanied with each report. It signifies a specific type of bug assigned by the tester in the company.
Reports with the same label denote they are duplicates of each other.
In this sense, we treat the pair of reports with the same label as duplicates, while the pair of reports with different labels as non-duplicates.

Table \ref{tab:projects} also presents the number and percentage of reports which have duplicates (i.e., \textit{Num dup.} and \textit{Pert. dup.}), the number of total pairs (calculated by $\mathit{num\_report * (num\_report-1) / 2}$), the number and percentage of duplicate pairs (i.e., \textit{Num dup. pairs} and \textit{Pert. dup. pairs}).

\begin{table}[!ht]
\scriptsize
\caption{\textbf{Projects under investigation}}
\label{tab:projects}
\centering\scalebox{0.9}{
\begin{tabular}{p{0.4cm}|p{1cm}|p{0.5cm}|p{0.5cm}|p{0.55cm}|p{0.5cm}|p{0.55cm}|p{0.65cm}|p{0.5cm}|p{0.5cm}}
\hline
&  \textbf{Domain} & \textbf{\# report} & \textbf{Num scr.} & \textbf{Pert. scr.} & \textbf{Num dup.} & \textbf{Pert. dup.} & \textbf{Total pairs} & \textbf{Num dup. pairs}  & \textbf{Pert. dup. pairs} \\
\hline
\textbf{P1} & Music  & 213 & 188 & 88\% & 208 & 97\% &  22578 &  7187    & 31\% \\
\hline
\textbf{P2} &  Weather & 215 & 200 & 93\% & 168 &  78\% &  22578 &  1549  & 7\%  \\
\hline
\textbf{P3} &  Beauty & 230 & 216 &  94\% & 214 &  93\% &  26335  &  5682  & 22\%  \\
\hline
\textbf{P4} & News  & 243 & 236 & 97\% & 210  & 86\% & 29403  &  5203 & 18\%  \\
\hline
\textbf{P5} & Browser  & 252 & 237 & 95\%  & 190  & 75\% & 31626 &  4347    & 14\% \\
\hline
\textbf{P6} & Medical  & 271 & 255 & 94\% & 207 &  76\% & 36585 &  1165    & 3\%  \\
\hline
\textbf{P7} & Safety  & 282 & 270 & 96\%  & 249 &  87\% & 40186 &  9358   & 23\%   \\
\hline
\textbf{P8} & Education  & 284 & 278 & 98\%  &233 & 82\%  & 40186 &  1753   & 4\%  \\
\hline
\textbf{P9} & Health  & 317 & 307 & 97\% & 246 & 77\% & 50086  &  1344  & 3\% \\
\hline
\textbf{P10} &  Language &  344 & 317 &  97\% & 236  & 68\% &  58996 &  1064   & 2\% \\
\hline
\textbf{P11} &  Sport &  462 & 425 & 93\% & 391  & 84\% &  106491  &  2381  & 2\%  \\
\hline
\textbf{P12} &  Efficiency &  576 & 547 & 95\% & 490  & 85\%  &  165600 &  17261  & 11\% \\
\hline
\multicolumn{2}{c|}{\textbf{Summary}}  & {\reports} & & 94\% & & 82\% &  &  & 12\%  \\
\hline
\end{tabular}
}
\end{table}

To verify the validity of these stored labels, we additionally conduct the random sampling and relabeling.
In detail, we randomly select 4 projects, and sample 30\% of crowdtesting reports from each selected project.
A tester from the company is asked to relabel the duplicate reports, without knowing the stored labels.
We then compare the difference between the stored duplicate results and the new labeled duplicate results.
The percentage of difference for each project is all below 4\%.
Therefore, we believe the ground truth labels are relatively trustworthy.

For \textbf{training the word embedding model}, we use another textual dataset. 
In detail, we crawl the textual description of crowdtesting reports and task requirements of 500 crowdtesting projects from the experimental platform.
The reason why we use this dataset is that previous studies have revealed that to train an effective word embedding model, a domain-specific dataset with large size is preferred~\cite{duplicate_embedding,duplicate_linkable_knowledge_stackoverflow}.
The size of our training dataset is 520M.

\subsection{Baselines}
\label{subsec:experiment_baseline}

To explore the performance of our proposed {\tool}, we compare it with three state-of-the-art and typical baseline approaches.
Note that, since there is no approach designed for duplicate \textit{crowdtesting report} detection, we choose the approaches for duplicate \textit{bug report} detection as our baselines.
Section \ref{subsec:related_dupIssue} has presented why we choose these three baselines.

\textbf{Information retrieval with word embedding (IR-EM)} \cite{duplicate_embedding}: It is the state-of-the-art technique for duplicate bug report detection.
This approach first builds TF-IDF vector and word embedding vector and calculates two similarity scores based on them respectively. 
Meanwhile, it calculates a third similarity score based on bug product field and component field. 
Finally, it combines the three similarity scores into one final score and makes similar bug recommendation with it.
% combines information retrieval and word embedding techniques, and takes the product and component information into consideration.

\textbf{Similarity based on bug components and descriptions (NextBug)} \cite{duplicate_SANER_empirical}: It is another state-of-the-art similarity-based approach for duplicate bug report detection.
This approach first checks whether two reports have the same bug component field, if yes, processes the reports with standard information retrieval technique, calculates the cosine similarity of the reports, and ranks the reports with the similarity value. 
% utilizes the similarity in bug components and bug descriptions to detect duplicate reports.

Note that, for these two baselines, because the crowdtesting reports do not have the product or component fields, we use the most similar field, i.e., \textit{test task id} for substitution.

\textbf{Information retrieval with topic modeling (DBTM)} \cite{duplicate_topic_model}:
It is the commonly-used technique for detecting duplicate bug reports.
DBTM supposes a report as a textual document describing one or more technical issues,
and duplicate reports as the documents describing the same technical issues.
It then utilizes term-based and topic-based features to detect duplicates.

\subsection{Experimental Setup}
\label{subsec:experiment_setup}

This section illustrates the experimental setup for answering each research question. 

\textbf{For answering RQ1}, we compare {\tool} with three state-of-the-art approaches (see details in Section \ref{subsec:experiment_baseline}) to investigate whether and to what extent it improves over prior work. 

\textbf{For answering RQ2}, we employ two additional experiments, i.e., \textit{onlyText} and \textit{onlyImage}, to investigate whether both screenshots and textual descriptions are necessary in duplicate reports detection.
In detail, \textit{onlyText} denotes ranking the reports only based on the textual similarity, while \textit{onlyImage} denotes ranking the reports only based on the screenshot similarity.

\textbf{For answering RQ3}, we use another four experiments,  i.e., \textit{noTF}, \textit{noEmb}, 
\textit{noClr}, and \textit{noStrc}, to investigate the relative effect of the four types of features utilized in duplicate detection. 
Each experiment denotes conducting the duplicate detection by removing one specific type of feature.
For example, \textit{noTF} denotes applying other three features except TF-IDF (i.e., only use word embedding, image color and image structure feature) for duplicate detection. 

% \textbf{For answering RQ4}, we design three new combination manners, i.e., \textit{addCmb}, \textit{multiplyCmb}, and \textit{textFirst}, to investigate whether other combination manner could achieve a better performance. 
% In detail, \textit{addCmb} denotes adding screenshot similarity and textual similarity as one similarity value and ranking the reports based on it, which is a straight-forward manner. 
% \textit{multiplyCmb} denotes multiplying the screenshot similarity with textual similarity as one similarity value and ranking the reports based on it, which is borrowed from \cite{duplicate_embedding}.
% \textit{textFirst} denotes treating the reports with high textual similarity as the first class and ranking them with the screenshot similarity (the second class is treated as {\tool} does).

For all these experiments, we employ the commonly-used leave-one-out cross validation~\cite{algorithm_nb_lr}. 
In detail, we use one crowdtesting project as the testing dataset to evaluate the performance of duplicate detection, and use the remaining crowdtesting projects as the training dataset to determine the optimal parameter value (see detail in Section \ref{subsec:experiment_parameter}).

\subsection{Evaluation Metrics}
\label{subsec:experiment_metric}

We use three evaluation metrics, i.e., recall@k, mean average precision (MAP), and mean reciprocal rank (MRR),
to evaluate the performance of duplicate detection.
These metrics are commonly-used to evaluate the duplicate detection approaches (see Table \ref{tab:relatedWork} for details).

% To briefly introduce these metrics, we first give some notations.
Given a query report $q$, its ground truth duplicate reports set $G(q)$, and the top-k recommended duplicate reports list produced by duplicate detection approach $R(q)$.

\textbf{Recall@k} checks whether a top-k recommendation is useful. 
The definition of $recall@k$ for a query bug $q$ is as follows: 

\begin{equation}
recall@k = 
\left\{
             \begin{array}{lr}
             1, \ if \ \mathit{G(q) \cap R(q)} \neq \varnothing \\
             0, \ Otherwise \\
             \end{array}
\right.
\end{equation}

According to the formula, if there is at least one ground truth duplicate report in the top-k recommendation, the top-k recommendation is useful for the query report $q$.
Given a set of query reports, we compute the proportion of useful top-k recommendations by averaging the $recall@k$ of all query reports to get an overall $recall@k$.
As previous approaches, we set $k$ as 1, 5, and 10 to obtain the performance.

\textbf{MAP} (Mean Average Precision) is defined as the mean of the
Average Precision (AP) values obtained for all the evaluation queries.
The AP of a single query $q$ is calculated as follows:

\begin{equation}
\label{eq-MAP}
AP(q) = \sum_{n=1}^{|G(q)|}\frac{Precision@k(q)}{|G(q)|}
\end{equation}

In the above formula, $Precision@k(q)$ is the retrieval precision over the top-k reports in the ranked list, i.e., the ratio of ground truth duplicate reports of the query report $q$ in the top-k recommendation:

\begin{equation}
\label{eq-MAP}
Precision@k(q) = \frac{\# \ ground \ truth \ in \ top \ k }{n}
\end{equation}

\textbf{MRR} (Mean Reciprocal Rank) is defined as the mean of the Reciprocal Rank (RR) values obtained for all the evaluation queries.
RR of a single query $q$ is the multiplicative inverse of the rank of first correct recommendation $first_{q}$ (i.e., first ground truth duplicate report in the recommendation list):

\begin{equation}
\label{eq-MRR}
RR(q) = \frac{1}{first_{q}}
\end{equation}

In addition, for each evaluation metric, we obtain the \textit{Improvement} of {\tool} compared with other approaches (e.g., the baseline). 
Taken metric \textit{MRR} and approach \textit{DBTM} as an example, \textit{Improvement} is calculated as follows: 

\begin{equation}
\label{eq-MRR}
Improvement = \frac{\mathit{MRR \ of \ {\tool}} - \mathit{MRR \ of \ DBTM}}{\mathit{\ MRR \ of \ DBTM}}
\end{equation}

To further demonstrate the superiority of our proposed approach, we perform the Mann–Whitney U test between our proposed {\tool} and other approaches (i.e., the baseline).
We obtain the \textbf{p-value} to demonstrate the significance of the test, and the \textbf{Cliff's delta} to demonstrate the effect size of the test. 
Mann–Whitney U test is often employed to check whether the difference in two data groups is statistically significant (which corresponds to a p-value of less than 0.05) or not. 
We use one-tailed Mann–Whitney U test with the following hypotheses:

\textit{H0: Performance produced by other approach (e.g., the baseline) is no smaller than the performance produced by {\tool}.}

\textit{H1: Performance produced by other approach (e.g., the baseline) is smalller than the performance produced by {\tool}.}

We include the Bonferroni correction to counteract the impact of multiple hypothesis tests. 

Cliff's delta is often used to check if the difference in two data groups are substantial. 
The range of Cliff's delta is [-1, 1], where -1 or 1 means all values in one group are smaller or larger than those of the other group, and 0 means the data in the two groups is similar. 
The mappings between Cliff's delta and effectiveness levels are shown below.

\begin{table}[!ht]
\scriptsize
\caption{\textbf{Mappings of Cliff's delta to their interpretations~\cite{cliff2014ordinal}}}
\label{tab:delta}
\centering{
\begin{tabular}{p{3cm}|p{3cm}}
\hline
Cliff's delta & Interpretation \\
\hline
-1 \textless= Cliff's delta \textless 0.147 & Negligible \\
\hline
0.147 \textless= Cliff's delta \textless 0.33 & Small \\
\hline
0.33 \textless= Cliff's delta \textless 0.474 & Medium \\
\hline
0.474 \textless= Cliff's delta \textless 1 & Large \\
\hline
\end{tabular}
}
\end{table}

Note that, we use all results from all query reports to compute p-value and Cliff's delta. 
For each query report, we have one value for other approach (i.e., the baseline) and another value for our proposed approach {\tool}. 
By computing the p-value and Cliff's delta, the extent of which our approach improves over other method can be more rigorously assessed.

\subsection{Parameter Setting}
\label{subsec:experiment_parameter}

\begin{figure}[ht!]
\centering
\includegraphics[width=7.2cm]{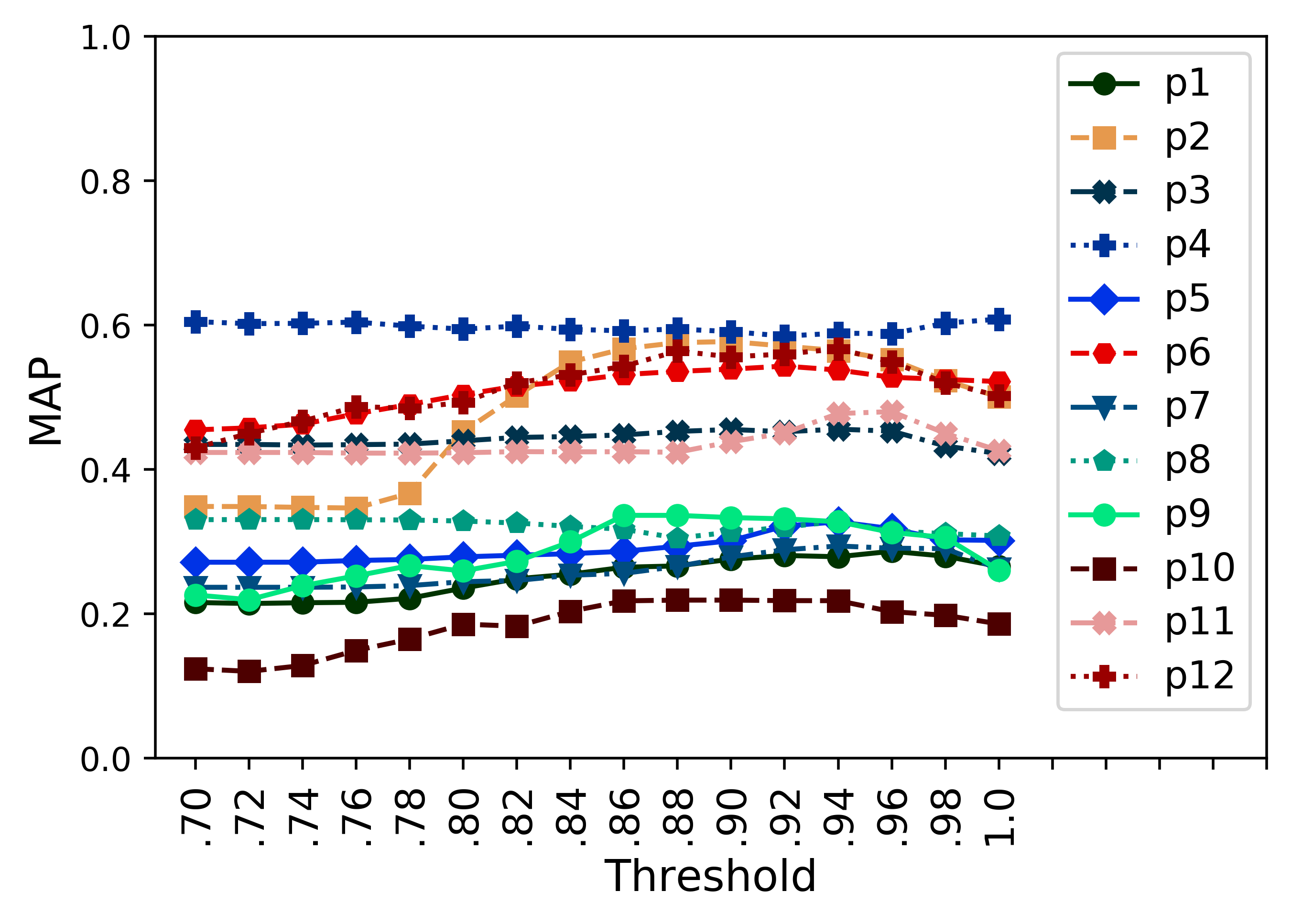}
\caption{Influence of parameter \textit{thres} on duplicate detection performance}
\label{fig:para}
\end{figure}

{\tool} has a parameter, i.e., \textit{thres}, to determine the  reports of first class (Section \ref{subsec:approach_detection}). 
Figure \ref{fig:para} presents how the duplicate detection performance is influenced by different parameter values. 
Note that, we only use \textit{MAP}, which is obtained considering the whole recommendation list of duplicates, to investigate the parameter's influence.  
Another note is that, we have experimented with \textit{thres} from 0.1 to 1.0, and due to space limit, we only present the results with better performance. 
We can easily observe that, almost for all our experimental projects, with the increase of \textit{thres}, the duplicate detection performance would first increase, reach a peak, and then decrease.

In our evaluation, we tune the optimal parameter value based on the training dataset (see Section \ref{subsec:experiment_setup}) and apply it in the testing dataset to evaluate the performance of duplicate detection. 
In detail, for each parameter value, we first compute the average \textit{MAP} across all the crowdtesting projects in the training dataset and treat the parameter value under which the largest average \textit{MAP} is achieved as the optimal \textit{thres}.
In this way, the tuned optimal parameter value is 0.94 for projects P1 - P10, and 0.92 for projects P11 and P12. 
For other experiments in our evaluation (i.e., \textit{noClr}), we use the same method to tune the optimal parameter value.

\section{Experiment Design}
\label{sec:experiment}

\subsection{Research Questions}
\label{subsec:experiment_rq}

Our evaluation addresses the following research questions:

\begin{itemize}
\item \textbf{RQ1 Effectiveness}: How effective is {\tool} in detecting duplicate crowdtesting reports?
\end{itemize}

RQ1 aims at evaluating the effectiveness of {\tool} in duplicate reports detection. 
We also compare {\tool} with the state-of-the-art approaches (see details in Section \ref{subsec:experiment_baseline}) to investigate whether and to what extent it improves over prior work. 

\begin{itemize}
\item \textbf{RQ2 Necessity}: Are both screenshots and textual descriptions necessary in detecting duplicate crowdtesting reports?
\end{itemize}

This paper proposes to utilize both screenshots and textual descriptions in duplicate detection. RQ2 is to investigate whether both of them are necessary. 
We employ two additional experiments to investigate it. 
See details in Section \ref{subsec:experiment_setup}.
% We employ two additional experiments, i.e., \textit{onlyText} and \textit{onlyImage}, to investigate this research question. 
% In detail, \textit{onlyText} denotes ranking the reports only based on the textual similarity (i.e., TF-IDF  similarity + word embedding similarity), while \textit{onlyImage} is based on the screenshot similarity (i.e., color similarity + structure similarity).

\begin{itemize}
\item \textbf{RQ3 Replaceability}: What is the relative effect of the four types of features (i.e., TF-IDF, word embedding, image color, and image structure) in detecting duplicate crowdtesting reports?
\end{itemize}

{\tool} employs four types of features to characterize the screenshots (i.e., image color and structure) and the textual descriptions (i.e., TF-IDF and word embedding). 
RQ3 is to investigate the relative effect of these features.
We use another four experiments for investigating this RQ, with details in Section \ref{subsec:experiment_setup}.
% We use another four experiments, i.e., \textit{noTF}, \textit{noEmb}, \textit{noClr} and \textit{noStrc}, to investigate this research question. 
% For example, \textit{noTF} denotes applying other three features except TF-IDF (i.e., only use embedding, image color and image structure feature) to conduct the duplicate detection. 

% \begin{itemize}
% \item \textbf{RQ4 Alternative}: Can other combination manner of screenshot similarity and textual similarity outperform {\tool} in detecting duplicate crowdtesting reports?
% \end{itemize}

% The motivating examples in Section \ref{subsec:background_motivation} indicate that
% the role of screenshots is mainly to demonstrate the context-related information, while the role of textual descriptions is to provide detailed illustration of the reported problem.  
% This is why our proposed approach {\tool} first uses screenshot similarity to filter the reports to first class and conducts the ranking separately. 

% However, one may still argue that other combination manners could achieve better performance.  
% RQ4 aims at investigating whether other combination manners could outperform {\tool} in duplicate detection. 
% We design three new combination manners to investigate this RQ. Details are shown in Section \ref{subsec:experiment_setup}.

\subsection{Experimental Dataset}
\label{subsec:experiment_data}

We mentioned that our experiment is based on crowdtesting reports from the repositories of {\company} crowdtesting platform.
We collect all crowdtesting projects closed between June 1st 2017 and June 10th 2017.
There are totally {\projects} crowdtesting projects. 

Table \ref{tab:projects} presents the detailed information of the projects with the application domain, the number of reports (i.e., \textit{\# report}), the number and percentage of reports which have screenshots (i.e., \textit{Num scr.} and \textit{Pert. scr.}).

There is a label accompanied with each report. It signifies a specific type of bug assigned by the tester in the company.
Reports with the same label denote they are duplicates of each other.
In this sense, we treat the pair of reports with the same label as duplicates, while the pair of reports with different labels as non-duplicates.

Table \ref{tab:projects} also presents the number and percentage of reports which have duplicates (i.e., \textit{Num dup.} and \textit{Pert. dup.}), the number of total pairs (calculated by $\mathit{num\_report * (num\_report-1) / 2}$), the number and percentage of duplicate pairs (i.e., \textit{Num dup. pairs} and \textit{Pert. dup. pairs}).

\begin{table}[!ht]
\scriptsize
\caption{\textbf{Projects under investigation}}
\label{tab:projects}
\centering\scalebox{0.9}{
\begin{tabular}{p{0.4cm}|p{1cm}|p{0.5cm}|p{0.5cm}|p{0.55cm}|p{0.5cm}|p{0.55cm}|p{0.65cm}|p{0.5cm}|p{0.5cm}}
\hline
&  \textbf{Domain} & \textbf{\# report} & \textbf{Num scr.} & \textbf{Pert. scr.} & \textbf{Num dup.} & \textbf{Pert. dup.} & \textbf{Total pairs} & \textbf{Num dup. pairs}  & \textbf{Pert. dup. pairs} \\
\hline
\textbf{P1} & Music  & 213 & 188 & 88\% & 208 & 97\% &  22578 &  7187    & 31\% \\
\hline
\textbf{P2} &  Weather & 215 & 200 & 93\% & 168 &  78\% &  22578 &  1549  & 7\%  \\
\hline
\textbf{P3} &  Beauty & 230 & 216 &  94\% & 214 &  93\% &  26335  &  5682  & 22\%  \\
\hline
\textbf{P4} & News  & 243 & 236 & 97\% & 210  & 86\% & 29403  &  5203 & 18\%  \\
\hline
\textbf{P5} & Browser  & 252 & 237 & 95\%  & 190  & 75\% & 31626 &  4347    & 14\% \\
\hline
\textbf{P6} & Medical  & 271 & 255 & 94\% & 207 &  76\% & 36585 &  1165    & 3\%  \\
\hline
\textbf{P7} & Safety  & 282 & 270 & 96\%  & 249 &  87\% & 40186 &  9358   & 23\%   \\
\hline
\textbf{P8} & Education  & 284 & 278 & 98\%  &233 & 82\%  & 40186 &  1753   & 4\%  \\
\hline
\textbf{P9} & Health  & 317 & 307 & 97\% & 246 & 77\% & 50086  &  1344  & 3\% \\
\hline
\textbf{P10} &  Language &  344 & 317 &  97\% & 236  & 68\% &  58996 &  1064   & 2\% \\
\hline
\textbf{P11} &  Sport &  462 & 425 & 93\% & 391  & 84\% &  106491  &  2381  & 2\%  \\
\hline
\textbf{P12} &  Efficiency &  576 & 547 & 95\% & 490  & 85\%  &  165600 &  17261  & 11\% \\
\hline
\multicolumn{2}{c|}{\textbf{Summary}}  & {\reports} & & 94\% & & 82\% &  &  & 12\%  \\
\hline
\end{tabular}
}
\end{table}

To verify the validity of these stored labels, we additionally conduct the random sampling and relabeling.
In detail, we randomly select 4 projects, and sample 30\% of crowdtesting reports from each selected project.
A tester from the company is asked to relabel the duplicate reports, without knowing the stored labels.
We then compare the difference between the stored duplicate results and the new labeled duplicate results.
The percentage of difference for each project is all below 4\%.
Therefore, we believe the ground truth labels are relatively trustworthy.

For \textbf{training the word embedding model}, we use another textual dataset. 
In detail, we crawl the textual description of crowdtesting reports and task requirements of 500 crowdtesting projects from the experimental platform.
The reason why we use this dataset is that previous studies have revealed that to train an effective word embedding model, a domain-specific dataset with large size is preferred~\cite{duplicate_embedding,duplicate_linkable_knowledge_stackoverflow}.
The size of our training dataset is 520M.

\subsection{Baselines}
\label{subsec:experiment_baseline}

To explore the performance of our proposed {\tool}, we compare it with three state-of-the-art and typical baseline approaches.
Note that, since there is no approach designed for duplicate \textit{crowdtesting report} detection, we choose the approaches for duplicate \textit{bug report} detection as our baselines.
Section \ref{subsec:related_dupIssue} has presented why we choose these three baselines.

\textbf{Information retrieval with word embedding (IR-EM)} \cite{duplicate_embedding}: It is the state-of-the-art technique for duplicate bug report detection.
This approach first builds TF-IDF vector and word embedding vector and calculates two similarity scores based on them respectively. 
Meanwhile, it calculates a third similarity score based on bug product field and component field. 
Finally, it combines the three similarity scores into one final score and makes similar bug recommendation with it.
% combines information retrieval and word embedding techniques, and takes the product and component information into consideration.

\textbf{Similarity based on bug components and descriptions (NextBug)} \cite{duplicate_SANER_empirical}: It is another state-of-the-art similarity-based approach for duplicate bug report detection.
This approach first checks whether two reports have the same bug component field, if yes, processes the reports with standard information retrieval technique, calculates the cosine similarity of the reports, and ranks the reports with the similarity value. 
% utilizes the similarity in bug components and bug descriptions to detect duplicate reports.

Note that, for these two baselines, because the crowdtesting reports do not have the product or component fields, we use the most similar field, i.e., \textit{test task id} for substitution.

\textbf{Information retrieval with topic modeling (DBTM)} \cite{duplicate_topic_model}:
It is the commonly-used technique for detecting duplicate bug reports.
DBTM supposes a report as a textual document describing one or more technical issues,
and duplicate reports as the documents describing the same technical issues.
It then utilizes term-based and topic-based features to detect duplicates.

\subsection{Experimental Setup}
\label{subsec:experiment_setup}

This section illustrates the experimental setup for answering each research question. 

\textbf{For answering RQ1}, we compare {\tool} with three state-of-the-art approaches (see details in Section \ref{subsec:experiment_baseline}) to investigate whether and to what extent it improves over prior work. 

\textbf{For answering RQ2}, we employ two additional experiments, i.e., \textit{onlyText} and \textit{onlyImage}, to investigate whether both screenshots and textual descriptions are necessary in duplicate reports detection.
In detail, \textit{onlyText} denotes ranking the reports only based on the textual similarity, while \textit{onlyImage} denotes ranking the reports only based on the screenshot similarity.

\textbf{For answering RQ3}, we use another four experiments,  i.e., \textit{noTF}, \textit{noEmb}, 
\textit{noClr}, and \textit{noStrc}, to investigate the relative effect of the four types of features utilized in duplicate detection. 
Each experiment denotes conducting the duplicate detection by removing one specific type of feature.
For example, \textit{noTF} denotes applying other three features except TF-IDF (i.e., only use word embedding, image color and image structure feature) for duplicate detection. 

% \textbf{For answering RQ4}, we design three new combination manners, i.e., \textit{addCmb}, \textit{multiplyCmb}, and \textit{textFirst}, to investigate whether other combination manner could achieve a better performance. 
% In detail, \textit{addCmb} denotes adding screenshot similarity and textual similarity as one similarity value and ranking the reports based on it, which is a straight-forward manner. 
% \textit{multiplyCmb} denotes multiplying the screenshot similarity with textual similarity as one similarity value and ranking the reports based on it, which is borrowed from \cite{duplicate_embedding}.
% \textit{textFirst} denotes treating the reports with high textual similarity as the first class and ranking them with the screenshot similarity (the second class is treated as {\tool} does).

For all these experiments, we employ the commonly-used leave-one-out cross validation~\cite{algorithm_nb_lr}. 
In detail, we use one crowdtesting project as the testing dataset to evaluate the performance of duplicate detection, and use the remaining crowdtesting projects as the training dataset to determine the optimal parameter value (see detail in Section \ref{subsec:experiment_parameter}).

\subsection{Evaluation Metrics}
\label{subsec:experiment_metric}

We use three evaluation metrics, i.e., recall@k, mean average precision (MAP), and mean reciprocal rank (MRR),
to evaluate the performance of duplicate detection.
These metrics are commonly-used to evaluate the duplicate detection approaches (see Table \ref{tab:relatedWork} for details).

% To briefly introduce these metrics, we first give some notations.
Given a query report $q$, its ground truth duplicate reports set $G(q)$, and the top-k recommended duplicate reports list produced by duplicate detection approach $R(q)$.

\textbf{Recall@k} checks whether a top-k recommendation is useful. 
The definition of $recall@k$ for a query bug $q$ is as follows: 

\begin{equation}
recall@k = 
\left\{
             \begin{array}{lr}
             1, \ if \ \mathit{G(q) \cap R(q)} \neq \varnothing \\
             0, \ Otherwise \\
             \end{array}
\right.
\end{equation}

According to the formula, if there is at least one ground truth duplicate report in the top-k recommendation, the top-k recommendation is useful for the query report $q$.
Given a set of query reports, we compute the proportion of useful top-k recommendations by averaging the $recall@k$ of all query reports to get an overall $recall@k$.
As previous approaches, we set $k$ as 1, 5, and 10 to obtain the performance.

\textbf{MAP} (Mean Average Precision) is defined as the mean of the
Average Precision (AP) values obtained for all the evaluation queries.
The AP of a single query $q$ is calculated as follows:

\begin{equation}
\label{eq-MAP}
AP(q) = \sum_{n=1}^{|G(q)|}\frac{Precision@k(q)}{|G(q)|}
\end{equation}

In the above formula, $Precision@k(q)$ is the retrieval precision over the top-k reports in the ranked list, i.e., the ratio of ground truth duplicate reports of the query report $q$ in the top-k recommendation:

\begin{equation}
\label{eq-MAP}
Precision@k(q) = \frac{\# \ ground \ truth \ in \ top \ k }{n}
\end{equation}

\textbf{MRR} (Mean Reciprocal Rank) is defined as the mean of the Reciprocal Rank (RR) values obtained for all the evaluation queries.
RR of a single query $q$ is the multiplicative inverse of the rank of first correct recommendation $first_{q}$ (i.e., first ground truth duplicate report in the recommendation list):

\begin{equation}
\label{eq-MRR}
RR(q) = \frac{1}{first_{q}}
\end{equation}

In addition, for each evaluation metric, we obtain the \textit{Improvement} of {\tool} compared with other approaches (e.g., the baseline). 
Taken metric \textit{MRR} and approach \textit{DBTM} as an example, \textit{Improvement} is calculated as follows: 

\begin{equation}
\label{eq-MRR}
Improvement = \frac{\mathit{MRR \ of \ {\tool}} - \mathit{MRR \ of \ DBTM}}{\mathit{\ MRR \ of \ DBTM}}
\end{equation}

To further demonstrate the superiority of our proposed approach, we perform the Mann–Whitney U test between our proposed {\tool} and other approaches (i.e., the baseline).
We obtain the \textbf{p-value} to demonstrate the significance of the test, and the \textbf{Cliff's delta} to demonstrate the effect size of the test. 
Mann–Whitney U test is often employed to check whether the difference in two data groups is statistically significant (which corresponds to a p-value of less than 0.05) or not. 
We use one-tailed Mann–Whitney U test with the following hypotheses:

\textit{H0: Performance produced by other approach (e.g., the baseline) is no smaller than the performance produced by {\tool}.}

\textit{H1: Performance produced by other approach (e.g., the baseline) is smalller than the performance produced by {\tool}.}

We include the Bonferroni correction to counteract the impact of multiple hypothesis tests. 

Cliff's delta is often used to check if the difference in two data groups are substantial. 
The range of Cliff's delta is [-1, 1], where -1 or 1 means all values in one group are smaller or larger than those of the other group, and 0 means the data in the two groups is similar. 
The mappings between Cliff's delta and effectiveness levels are shown below.

\begin{table}[!ht]
\scriptsize
\caption{\textbf{Mappings of Cliff's delta to their interpretations~\cite{cliff2014ordinal}}}
\label{tab:delta}
\centering{
\begin{tabular}{p{3cm}|p{3cm}}
\hline
Cliff's delta & Interpretation \\
\hline
-1 \textless= Cliff's delta \textless 0.147 & Negligible \\
\hline
0.147 \textless= Cliff's delta \textless 0.33 & Small \\
\hline
0.33 \textless= Cliff's delta \textless 0.474 & Medium \\
\hline
0.474 \textless= Cliff's delta \textless 1 & Large \\
\hline
\end{tabular}
}
\end{table}

Note that, we use all results from all query reports to compute p-value and Cliff's delta. 
For each query report, we have one value for other approach (i.e., the baseline) and another value for our proposed approach {\tool}. 
By computing the p-value and Cliff's delta, the extent of which our approach improves over other method can be more rigorously assessed.

\subsection{Parameter Setting}
\label{subsec:experiment_parameter}

\begin{figure}[ht!]
\centering
\includegraphics[width=7.2cm]{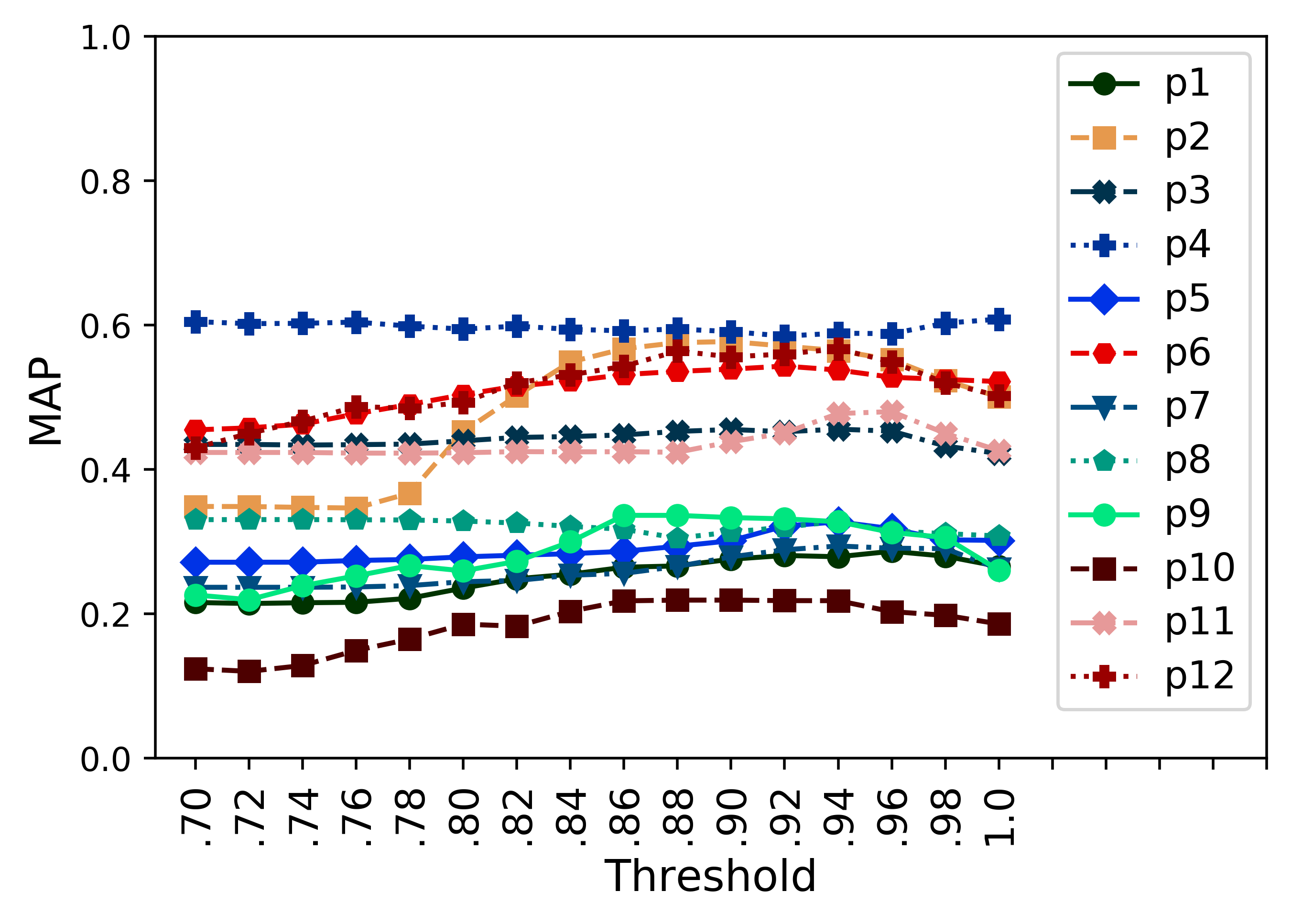}
\caption{Influence of parameter \textit{thres} on duplicate detection performance}
\label{fig:para}
\end{figure}

{\tool} has a parameter, i.e., \textit{thres}, to determine the  reports of first class (Section \ref{subsec:approach_detection}). 
Figure \ref{fig:para} presents how the duplicate detection performance is influenced by different parameter values. 
Note that, we only use \textit{MAP}, which is obtained considering the whole recommendation list of duplicates, to investigate the parameter's influence.  
Another note is that, we have experimented with \textit{thres} from 0.1 to 1.0, and due to space limit, we only present the results with better performance. 
We can easily observe that, almost for all our experimental projects, with the increase of \textit{thres}, the duplicate detection performance would first increase, reach a peak, and then decrease.

In our evaluation, we tune the optimal parameter value based on the training dataset (see Section \ref{subsec:experiment_setup}) and apply it in the testing dataset to evaluate the performance of duplicate detection. 
In detail, for each parameter value, we first compute the average \textit{MAP} across all the crowdtesting projects in the training dataset and treat the parameter value under which the largest average \textit{MAP} is achieved as the optimal \textit{thres}.
In this way, the tuned optimal parameter value is 0.94 for projects P1 - P10, and 0.92 for projects P11 and P12. 
For other experiments in our evaluation (i.e., \textit{noClr}), we use the same method to tune the optimal parameter value.

\section{Results and Analysis}
\label{sec:result}

This section presents the results and analysis of the evaluation.

\subsection{Answering RQ1: Effectiveness}
\label{subsec:result_rq1}

\begin{table*}[!t]
\scriptsize
\caption{\textbf{Performance comparison between {\tool} and baselines (RQ1)}}
\label{tab:rq1}
\centering{
\vspace{0.05in}
\begin{tabular}{p{0.9cm}|p{0.5cm}|p{0.5cm}|p{0.5cm}|p{0.5cm}|p{1.4cm}|p{0.5cm}|p{0.5cm}|p{0.5cm}|p{0.5cm}|p{1.3cm}|p{0.5cm}|p{0.5cm}|p{0.5cm}|p{0.5cm}|p{1.3cm}}
\hline
& \textbf{SETU} & \textbf{IR-EM} & \textbf{Next   Bug} & \textbf{DB   TM} & \textbf{Improvement} & \textbf{SETU} & \textbf{IR-EM} & \textbf{Next   Bug} & \textbf{DB    TM} & \textbf{Improvement} &  \textbf{SETU} & \textbf{IR-EM} & \textbf{Next   Bug} & \textbf{DB   TM} & \textbf{Improvement} \\ \hline
& \multicolumn{5}{|c}{\textbf{P1}} & \multicolumn{5}{|c}{\textbf{P2}} & \multicolumn{5}{|c}{\textbf{P3}} \\ \hline
recall@1  & \textbf{0.792}  & 0.640 & 0.440 & 0.424 & \textit{23\% - 86\%}  & \textbf{0.649}  & 0.520 & 0.470 & 0.537 & \textit{20\% - 38\%}  & \textbf{0.715}  & 0.575 & 0.487 & 0.535 & \textit{24\% - 46\%} \\ \hline
recall@5  & \textbf{0.872}  & 0.720 & 0.640 & 0.700 & \textit{21\% - 36\%}  & \textbf{0.836}  & 0.594 & 0.544 & 0.601 & \textit{39\% - 53\%}  & \textbf{0.894}  & 0.654 & 0.504 & 0.602 & \textit{36\% - 77\%} \\ \hline
recall@10  & \textbf{0.944}  & 0.780 & 0.728 & 0.750 & \textit{21\% - 29\%}  & \textbf{0.875}  & 0.664 & 0.604 & 0.665 & \textit{31\% - 44\%}  & \textbf{0.915}  & 0.735 & 0.664 & 0.720 & \textit{24\% - 37\%} \\ \hline
MAP  & \textbf{0.280}  & 0.188 & 0.158 & 0.151 & \textit{48\% - 85\%}  & \textbf{0.570}  & 0.336 & 0.283 & 0.389 & \textit{46\% - 101\%}  & \textbf{0.452}  & 0.333 & 0.213 & 0.321 & \textit{35\% - 112\%} \\ \hline
MRR  & \textbf{0.831}  & 0.684 & 0.624 & 0.576 & \textit{21\% - 44\%}  & \textbf{0.736}  & 0.555 & 0.505 & 0.575 & \textit{28\% - 45\%}  & \textbf{0.794}  & 0.616 & 0.576 & 0.601 & \textit{28\% - 37\%} \\ \hline
& \multicolumn{5}{|c}{\textbf{P4}} & \multicolumn{5}{|c}{\textbf{P5}} & \multicolumn{5}{|c}{\textbf{P6}} \\ \hline
recall@1  & \textbf{0.729}  & 0.433 & 0.363 & 0.445 & \textit{63\% - 100\%}  & \textbf{0.553}  & 0.418 & 0.298 & 0.386 & \textit{32\% - 85\%}  & \textbf{0.722}  & 0.573 & 0.505 & 0.522 & \textit{26\% - 42\%} \\ \hline
recall@5  & \textbf{0.927}  & 0.660 & 0.460 & 0.689 & \textit{34\% - 101\%}  & \textbf{0.815}  & 0.675 & 0.575 & 0.615 & \textit{20\% - 41\%}  & \textbf{0.871}  & 0.669 & 0.609 & 0.632 & \textit{30\% - 43\%} \\ \hline
recall@10  & \textbf{0.958}  & 0.761 & 0.561 & 0.772 & \textit{24\% - 70\%}  & \textbf{0.922}  & 0.745 & 0.685 & 0.739 & \textit{23\% - 34\%}  & \textbf{0.915}  & 0.759 & 0.739 & 0.735 & \textit{20\% - 24\%} \\ \hline
MAP  & \textbf{0.584}  & 0.171 & 0.271 & 0.198 & \textit{115\% - 241\%}  & \textbf{0.288}  & 0.161 & 0.119 & 0.141 & \textit{78\% - 142\%}  & \textbf{0.543}  & 0.422 & 0.362 & 0.294 & \textit{28\% - 84\%} \\ \hline
MRR  & \textbf{0.815}  & 0.527 & 0.417 & 0.536 & \textit{52\% - 95\%}  & \textbf{0.663}  & 0.490 & 0.320 & 0.447 & \textit{35\% - 107\%}  & \textbf{0.792}  & 0.631 & 0.531 & 0.607 & \textit{25\% - 49\%} \\ \hline
& \multicolumn{5}{|c}{\textbf{P7}} & \multicolumn{5}{|c}{\textbf{P8}} & \multicolumn{5}{|c}{\textbf{P9}} \\ \hline
recall@1  & \textbf{0.542}  & 0.174 & 0.224 & 0.193 & \textit{141\% - 211\%}  & \textbf{0.647}  & 0.485 & 0.415 & 0.521 & \textit{24\% - 55\%}  & \textbf{0.549}  & 0.403 & 0.353 & 0.372 & \textit{36\% - 55\%} \\ \hline
recall@5  & \textbf{0.666}  & 0.341 & 0.300 & 0.335 & \textit{95\% - 122\%}  & \textbf{0.849}  & 0.667 & 0.583 & 0.605 & \textit{27\% - 45\%}  & \textbf{0.768}  & 0.568 & 0.528 & 0.555 & \textit{35\% - 45\%} \\ \hline
recall@10  & \textbf{0.693}  & 0.440 & 0.420 & 0.409 & \textit{57\% - 69\%}  & \textbf{0.877}  & 0.702 & 0.664 & 0.712 & \textit{23\% - 32\%}  & \textbf{0.811}  & 0.637 & 0.587 & 0.606 & \textit{27\% - 38\%} \\ \hline
MAP  & \textbf{0.259}  & 0.090 & 0.080 & 0.100 & \textit{159\% - 223\%}  & \textbf{0.321}  & 0.244 & 0.164 & 0.220 & \textit{31\% - 95\%}  & \textbf{0.307}  & 0.210 & 0.210 & 0.195 & \textit{46\% - 57\%} \\ \hline
MRR  & \textbf{0.565}  & 0.245 & 0.215 & 0.218 & \textit{130\% - 162\%}  & \textbf{0.738}  & 0.597 & 0.473 & 0.571 & \textit{23\% - 56\%}  & \textbf{0.644}  & 0.527 & 0.477 & 0.501 & \textit{22\% - 35\%} \\ \hline
& \multicolumn{5}{|c}{\textbf{P10}} & \multicolumn{5}{|c}{\textbf{P11}} & \multicolumn{5}{|c}{\textbf{P12}} \\ \hline
recall@1  & \textbf{0.440}  & 0.293 & 0.223 & 0.257 & \textit{50\% - 97\%}  & \textbf{0.773}  & 0.613 & 0.523 & 0.512 & \textit{26\% - 50\%}  & \textbf{0.719}  & 0.578 & 0.488 & 0.525 & \textit{24\% - 47\%} \\ \hline
recall@5  & \textbf{0.695}  & 0.421 & 0.321 & 0.387 & \textit{65\% - 116\%}  & \textbf{0.887}  & 0.729 & 0.659 & 0.662 & \textit{21\% - 34\%}  & \textbf{0.927}  & 0.724 & 0.684 & 0.702 & \textit{28\% - 35\%} \\ \hline
recall@10  & \textbf{0.726}  & 0.585 & 0.486 & 0.530 & \textit{24\% - 49\%}  & \textbf{0.924}  & 0.759 & 0.739 & 0.725 & \textit{21\% - 27\%}  & \textbf{0.948}  & 0.764 & 0.734 & 0.750 & \textit{24\% - 29\%} \\ \hline
MAP  & \textbf{0.219}  & 0.159 & 0.129 & 0.143 & \textit{37\% - 69\%}  & \textbf{0.450}  & 0.332 & 0.282 & 0.274 & \textit{35\% - 64\%}  & \textbf{0.564}  & 0.413 & 0.383 & 0.401 & \textit{36\% - 47\%} \\ \hline
MRR  & \textbf{0.552}  & 0.400 & 0.370 & 0.412 & \textit{33\% - 49\%}  & \textbf{0.828}  & 0.621 & 0.531 & 0.577 & \textit{33\% - 55\%}  & \textbf{0.805}  & 0.646 & 0.586 & 0.601 & \textit{24\% - 37\%} \\ \hline
\end{tabular}
}
\end{table*}

\begin{table*}[!t]
\scriptsize
\caption{\textbf{Results of Mann-Whitney U Test between {\tool} and baselines (RQ1)}}
\label{tab:rq1-test}
\centering{
\vspace{0.05in}
\begin{tabular}{p{0.9cm}|p{1.4cm}|p{1.4cm}|p{1.4cm}|p{1.4cm}|p{1.4cm}|p{1.4cm}|p{1.4cm}|p{1.4cm}|p{1.4cm}}
\hline
&  \textbf{SETU vs. IR-EM} & \textbf{SETU vs.   NextBug} & \textbf{SETU vs.   DBTM} &   \textbf{SETU vs.   IR-EM} & \textbf{SETU vs. NextBug} & \textbf{SETU vs.    DBTM}  &  \textbf{SETU vs.   IR-EM} & \textbf{SETU vs.   NextBug} & \textbf{SETU vs.   DBTM} \\ \hline
& \multicolumn{3}{|c}{\textbf{P1}} & \multicolumn{3}{|c}{\textbf{P2}} & \multicolumn{3}{|c}{\textbf{P3}} \\ \hline
recall@1  & 0.00 (0.76 L) & 0.00 (0.99 L) & 0.00 (0.99 L) & 0.00 (0.71 L) & 0.00 (0.88 L) & 0.00 (0.66 L) & 0.00 (0.64 L) & 0.00 (0.89 L) & 0.00 (0.80 L)\\ \hline
recall@5  & 0.00 (0.66 L) & 0.00 (0.89 L) & 0.00 (0.79 L) & 0.00 (0.91 L) & 0.00 (0.98 L) & 0.00 (0.91 L) & 0.00 (0.90 L) & 0.00 (0.99 L) & 0.00 (0.95 L)\\ \hline
recall@10  & 0.00 (0.73 L) & 0.00 (0.86 L) & 0.00 (0.77 L) & 0.00 (0.85 L) & 0.00 (0.91 L) & 0.00 (0.83 L) & 0.00 (0.74 L) & 0.00 (0.92 L) & 0.00 (0.81 L)\\ \hline
MAP  & 0.00 (0.81 L) & 0.00 (0.90 L) & 0.00 (0.93 L) & 0.00 (0.94 L) & 0.00 (0.97 L) & 0.00 (0.81 L) & 0.00 (0.64 L) & 0.00 (0.97 L) & 0.00 (0.77 L)\\ \hline
MRR  & 0.00 (0.66 L) & 0.00 (0.79 L) & 0.00 (0.93 L) & 0.00 (0.77 L) & 0.00 (0.90 L) & 0.00 (0.74 L) & 0.00 (0.78 L) & 0.00 (0.85 L) & 0.00 (0.80 L)\\ \hline
& \multicolumn{3}{|c}{\textbf{P4}} & \multicolumn{3}{|c}{\textbf{P5}} & \multicolumn{3}{|c}{\textbf{P6}} \\ \hline
recall@1  & 0.00 (0.98 L) & 0.00 (0.99 L) & 0.00 (0.97 L) & 0.00 (0.75 L) & 0.00 (0.97 L) & 0.00 (0.86 L) & 0.00 (0.72 L) & 0.00 (0.91 L) & 0.00 (0.86 L)\\ \hline
recall@5  & 0.00 (0.90 L) & 0.00 (1.0 L) & 0.00 (0.84 L) & 0.00 (0.67 L) & 0.00 (0.92 L) & 0.00 (0.80 L) & 0.00 (0.74 L) & 0.00 (0.90 L) & 0.00 (0.87 L)\\ \hline
recall@10  & 0.00 (0.91 L) & 0.00 (0.99 L) & 0.00 (0.75 L) & 0.00 (0.67 L) & 0.00 (0.87 L) & 0.00 (0.78 L) & 0.00 (0.62 L) & 0.00 (0.68 L) & 0.00 (0.77 L)\\ \hline
MAP  & 0.00 (1.0 L) & 0.00 (0.99 L) & 0.00 (1.0 L) & 0.00 (0.90 L) & 0.00 (0.98 L) & 0.00 (0.93 L) & 0.00 (0.64 L) & 0.00 (0.88 L) & 0.00 (0.97 L)\\ \hline
MRR  & 0.00 (0.95 L) & 0.00 (0.99 L) & 0.00 (0.93 L) & 0.00 (0.71 L) & 0.00 (0.98 L) & 0.00 (0.83 L) & 0.00 (0.74 L) & 0.00 (0.92 L) & 0.00 (0.84 L)\\ \hline
& \multicolumn{3}{|c}{\textbf{P7}} & \multicolumn{3}{|c}{\textbf{P8}} & \multicolumn{3}{|c}{\textbf{P9}} \\ \hline
recall@1  & 0.00 (0.96 L) & 0.00 (0.92 L) & 0.00 (0.96 L) & 0.00 (0.77 L) & 0.00 (0.92 L) & 0.00 (0.64 L) & 0.00 (0.76 L) & 0.00 (0.87 L) & 0.00 (0.86 L)\\ \hline
recall@5  & 0.00 (0.97 L) & 0.00 (0.99 L) & 0.00 (0.98 L) & 0.00 (0.84 L) & 0.00 (0.94 L) & 0.00 (0.92 L) & 0.00 (0.83 L) & 0.00 (0.89 L) & 0.00 (0.82 L)\\ \hline
recall@10  & 0.00 (0.96 L) & 0.00 (0.97 L) & 0.00 (0.98 L) & 0.00 (0.69 L) & 0.00 (0.80 L) & 0.00 (0.72 L) & 0.00 (0.79 L) & 0.00 (0.87 L) & 0.00 (0.85 L)\\ \hline
MAP  & 0.00 (0.97 L) & 0.00 (0.97 L) & 0.00 (0.96 L) & 0.00 (0.65 L) & 0.00 (0.95 L) & 0.00 (0.79 L) & 0.00 (0.72 L) & 0.00 (0.72 L) & 0.00 (0.79 L)\\ \hline
MRR  & 0.00 (0.99 L) & 0.00 (0.99 L) & 0.00 (0.98 L) & 0.00 (0.72 L) & 0.00 (0.96 L) & 0.00 (0.83 L) & 0.00 (0.62 L) & 0.00 (0.79 L) & 0.00 (0.76 L)\\ \hline
& \multicolumn{3}{|c}{\textbf{P10}} & \multicolumn{3}{|c}{\textbf{P11}} & \multicolumn{3}{|c}{\textbf{P12}} \\ \hline
recall@1  & 0.00 (0.90 L) & 0.00 (0.98 L) & 0.00 (0.94 L) & 0.00 (0.67 L) & 0.00 (0.94 L) & 0.00 (0.95 L) & 0.00 (0.62 L) & 0.00 (0.91 L) & 0.00 (0.82 L)\\ \hline
recall@5  & 0.00 (0.97 L) & 0.00 (0.99 L) & 0.00 (0.99 L) & 0.00 (0.61 L) & 0.00 (0.83 L) & 0.00 (0.83 L) & 0.00 (0.83 L) & 0.00 (0.92 L) & 0.00 (0.86 L)\\ \hline
recall@10  & 0.00 (0.63 L) & 0.00 (0.91 L) & 0.00 (0.86 L) & 0.00 (0.70 L) & 0.00 (0.74 L) & 0.00 (0.80 L) & 0.00 (0.77 L) & 0.00 (0.83 L) & 0.00 (0.79 L)\\ \hline
MAP  & 0.00 (0.63 L) & 0.00 (0.75 L) & 0.00 (0.72 L) & 0.00 (0.76 L) & 0.00 (0.90 L) & 0.00 (0.93 L) & 0.00 (0.78 L) & 0.00 (0.88 L) & 0.00 (0.85 L)\\ \hline
MRR  & 0.00 (0.77 L) & 0.00 (0.87 L) & 0.00 (0.76 L) & 0.00 (0.86 L) & 0.00 (0.96 L) & 0.00 (0.92 L) & 0.00 (0.73 L) & 0.00 (0.90 L) & 0.00 (0.87 L)\\ \hline

\end{tabular}
}
\scriptsize{Note that: The figures X (Y Z) respectively denotes p-value, Cliffs delta, and interpretation of Cliffs delta (i.e., Large (L), Medium (M), Small (S), and Negligible (N))}
\end{table*}

Table \ref{tab:rq1} presents the \textit{recall@1}, \textit{recall@5}, \textit{recall@10}, \textit{MAP}, and \textit{MRR} for each experimental project for {\tool} and three baselines.

We can see that, our approach {\tool} can achieve the highest performance in all experimental projects for all five evaluation metrics.
\textit{recall@1} is 0.44 to 0.79 across all experimental projects, denoting in 44\% to 79\% circumstances, our approach can find the duplicate report in the first recommendation. 
\textit{recall@5} is 0.66 to 0.92 across all experimental projects, denoting in 66\% to 92\% circumstances, our first five recommendations contain the duplicate report. 
\textit{MAP} is 0.21 to 0.58 across all experimental projects. 
The reported \textit{recall@1} in existing duplicate bug report detection approach is 0.16 to 0.67, and the reported MAP is 0.26 to 0.53 (see Table \ref{tab:relatedWork}).
Since our experiment is conducted on crowdtesting reports which is different from bug reports in open source projects, these figures are not comparable. 
However, the fact that these figures being the same order of magnitude proves the effectiveness of our approach.

Compared with the three baselines, {\tool} brings great improvement in all five evaluation metrics for all experimental projects. 
The improvement of \textit{recall@1} is 20\% to 211\% compared with the three baselines, while the improvement of MAP is 28\% to 241\% in all experimental projects.
Other evaluation metrics also undergo similar improvements.

We further conduct Mann-Whitney U Test for the five metrics between {\tool} and each baseline.
Table \ref{tab:rq1-test} shows the p-value, the Cliff’s delta, and the interpretation for these tests (see Section \ref{subsec:experiment_metric} for details).
We can see that, compared with the three baselines, {\tool} statistically significantly (i.e., p-value for all tests is less than 0.05) and substantially (i.e., Cliff's delta for all tests is large ) achieves a better performance in terms of all the evaluation metrics for all experimental projects. 
This further indicates the effectiveness and advantage of our approach.

\begin{table*}[!t]
\scriptsize
\caption{\textbf{Performance comparison between {\tool} and \textit{onlyText}, \textit{onlyImage} (RQ2)}}
\label{tab:rq2}
\centering\scalebox{0.9}{
\vspace{0.05in}
\begin{tabular}{p{0.85cm}|p{0.5cm}|p{0.5cm}|p{0.5cm}|p{1.25cm}|p{0.5cm}|p{0.5cm}|p{0.5cm}|p{1.25cm}|p{0.5cm}|p{0.5cm}|p{0.5cm}|p{1.35cm}|p{0.5cm}|p{0.5cm}|p{0.5cm}|p{1.35cm}}
\hline
& \textbf{SETU} & \textbf{only Text} & \textbf{only  Image} & \textbf{Improvement} &  \textbf{SETU} & \textbf{only  Text} & \textbf{only  Image} & \textbf{Improvement} &  \textbf{SETU} & \textbf{only Text} & \textbf{only  Image} & \textbf{Improvement} &  \textbf{SETU} & \textbf{only Text} & \textbf{only   Image} & \textbf{Improvement} \\ \hline
& \multicolumn{4}{|c}{\textbf{P1}} & \multicolumn{4}{|c}{\textbf{P2}} & \multicolumn{4}{|c}{\textbf{P3}} & \multicolumn{4}{|c}{\textbf{P4}} \\ \hline
recall@1  & \textbf{0.792}  & 0.640 & 0.560 & \textit{23\% - 41\%}  & \textbf{0.649}  & 0.520 & 0.505 & \textit{24\% - 28\%}  & \textbf{0.715}  & 0.550 & 0.457 & \textit{29\% - 56\%}  & \textbf{0.729}  & 0.433 & 0.397 & \textit{68\% - 83\%} \\ \hline
recall@5  & \textbf{0.872}  & 0.720 & 0.640 & \textit{21\% - 36\%}  & \textbf{0.836}  & 0.584 & 0.550 & \textit{43\% - 51\%}  & \textbf{0.894}  & 0.634 & 0.568 & \textit{41\% - 57\%}  & \textbf{0.927}  & 0.660 & 0.643 & \textit{40\% - 44\%} \\ \hline
recall@10  & \textbf{0.944}  & 0.780 & 0.664 & \textit{21\% - 42\%}  & \textbf{0.875}  & 0.646 & 0.601 & \textit{35\% - 45\%}  & \textbf{0.915}  & 0.715 & 0.647 & \textit{27\% - 41\%}  & \textbf{0.958}  & 0.761 & 0.725 & \textit{25\% - 32\%} \\ \hline
MAP  & \textbf{0.280}  & 0.188 & 0.165 & \textit{48\% - 69\%}  & \textbf{0.570}  & 0.326 & 0.257 & \textit{74\% - 121\%}  & \textbf{0.452}  & 0.313 & 0.283 & \textit{44\% - 59\%}  & \textbf{0.584}  & 0.171 & 0.178 & \textit{228\% - 241\%} \\ \hline
MRR  & \textbf{0.831}  & 0.684 & 0.600 & \textit{21\% - 38\%}  & \textbf{0.736}  & 0.555 & 0.422 & \textit{32\% - 74\%}  & \textbf{0.794}  & 0.596 & 0.512 & \textit{33\% - 55\%}  & \textbf{0.815}  & 0.527 & 0.467 & \textit{54\% - 74\%} \\ \hline
& \multicolumn{4}{|c}{\textbf{P5}} & \multicolumn{4}{|c}{\textbf{P6}} & \multicolumn{4}{|c}{\textbf{P7}} & \multicolumn{4}{|c}{\textbf{P8}} \\ \hline
recall@1  & \textbf{0.553}  & 0.418 & 0.363 & \textit{32\% - 52\%}  & \textbf{0.722}  & 0.553 & 0.614 & \textit{17\% - 30\%}  & \textbf{0.542}  & 0.174 & 0.184 & \textit{194\% - 211\%}  & \textbf{0.647}  & 0.485 & 0.342 & \textit{33\% - 89\%} \\ \hline
recall@5  & \textbf{0.815}  & 0.675 & 0.630 & \textit{20\% - 29\%}  & \textbf{0.871}  & 0.639 & 0.635 & \textit{36\% - 37\%}  & \textbf{0.666}  & 0.341 & 0.331 & \textit{95\% - 101\%}  & \textbf{0.849}  & 0.667 & 0.488 & \textit{27\% - 73\%} \\ \hline
recall@10  & \textbf{0.922}  & 0.745 & 0.660 & \textit{23\% - 39\%}  & \textbf{0.915}  & 0.749 & 0.689 & \textit{22\% - 32\%}  & \textbf{0.693}  & 0.440 & 0.420 & \textit{57\% - 64\%}  & \textbf{0.877}  & 0.702 & 0.511 & \textit{24\% - 71\%} \\ \hline
MAP  & \textbf{0.288}  & 0.161 & 0.147 & \textit{78\% - 95\%}  & \textbf{0.543}  & 0.402 & 0.367 & \textit{35\% - 47\%}  & \textbf{0.259}  & 0.090 & 0.080 & \textit{187\% - 223\%}  & \textbf{0.321}  & 0.244 & 0.212 & \textit{31\% - 51\%} \\ \hline
MRR  & \textbf{0.663}  & 0.490 & 0.463 & \textit{35\% - 43\%}  & \textbf{0.792}  & 0.611 & 0.569 & \textit{29\% - 39\%}  & \textbf{0.565}  & 0.245 & 0.235 & \textit{130\% - 140\%}  & \textbf{0.738}  & 0.597 & 0.407 & \textit{23\% - 81\%} \\ \hline
& \multicolumn{4}{|c}{\textbf{P9}} & \multicolumn{4}{|c}{\textbf{P10}} & \multicolumn{4}{|c}{\textbf{P11}} & \multicolumn{4}{|c}{\textbf{P12}} \\ \hline
recall@1  & \textbf{0.549}  & 0.383 & 0.138 & \textit{43\% - 297\%}  & \textbf{0.440}  & 0.293 & 0.105 & \textit{50\% - 319\%}  & \textbf{0.773}  & 0.593 & 0.555 & \textit{30\% - 39\%}  & \textbf{0.719}  & 0.538 & 0.438 & \textit{33\% - 64\%} \\ \hline
recall@5  & \textbf{0.768}  & 0.528 & 0.193 & \textit{45\% - 297\%}  & \textbf{0.695}  & 0.421 & 0.391 & \textit{65\% - 77\%}  & \textbf{0.887}  & 0.709 & 0.622 & \textit{25\% - 42\%}  & \textbf{0.927}  & 0.704 & 0.604 & \textit{31\% - 53\%} \\ \hline
recall@10  & \textbf{0.811}  & 0.617 & 0.339 & \textit{31\% - 139\%}  & \textbf{0.726}  & 0.585 & 0.565 & \textit{24\% - 28\%}  & \textbf{0.924}  & 0.729 & 0.655 & \textit{26\% - 41\%}  & \textbf{0.948}  & 0.764 & 0.664 & \textit{24\% - 42\%} \\ \hline
MAP  & \textbf{0.307}  & 0.180 & 0.047 & \textit{70\% - 553\%}  & \textbf{0.219}  & 0.159 & 0.096 & \textit{37\% - 128\%}  & \textbf{0.450}  & 0.312 & 0.321 & \textit{40\% - 44\%}  & \textbf{0.564}  & 0.403 & 0.303 & \textit{39\% - 86\%} \\ \hline
MRR  & \textbf{0.644}  & 0.477 & 0.121 & \textit{35\% - 432\%}  & \textbf{0.552}  & 0.400 & 0.229 & \textit{38\% - 141\%}  & \textbf{0.828}  & 0.601 & 0.593 & \textit{37\% - 39\%}  & \textbf{0.805}  & 0.606 & 0.506 & \textit{32\% - 59\%} \\ \hline
\end{tabular}
}
\end{table*}

\begin{table*}[!t]
\scriptsize
\caption{\textbf{Results of Mann-Whitney U Test between {\tool} and \textit{onlyText}, \textit{onlyImage} (RQ2)}}
\label{tab:rq2-test}
\centering{
\vspace{0.05in}
\begin{tabular}{p{0.9cm}|p{1.5cm}|p{1.5cm}|p{1.5cm}|p{1.5cm}|p{1.5cm}|p{1.5cm}|p{1.5cm}|p{1.5cm}}
\hline
&  \textbf{SETU vs. onlyText} & \textbf{SETU vs.  onlyImage} & \textbf{SETU vs.  onlyText} &   \textbf{SETU vs.   onlyImage} & \textbf{SETU vs. onlyText} & \textbf{SETU vs.  onlyImage} & \textbf{SETU vs.  onlyText} & \textbf{SETU vs.  onlyImage}  \\ \hline
& \multicolumn{2}{|c}{\textbf{P1}} & \multicolumn{2}{|c}{\textbf{P2}} & \multicolumn{2}{|c}{\textbf{P3}} & \multicolumn{2}{|c}{\textbf{P4}}\\ \hline
recall@1  & 0.00 (0.63 L) & 0.00 (0.89 L) & 0.00 (0.70 L) & 0.00 (0.78 L) & 0.00 (0.78 L) & 0.00 (0.94 L) & 0.00 (0.96 L) & 0.00 (0.97 L)\\ \hline
recall@5  & 0.00 (0.64 L) & 0.00 (0.78 L) & 0.00 (0.90 L) & 0.00 (0.93 L) & 0.00 (0.90 L) & 0.00 (0.98 L) & 0.00 (0.94 L) & 0.00 (0.93 L)\\ \hline
recall@10  & 0.00 (0.67 L) & 0.00 (0.90 L) & 0.00 (0.83 L) & 0.00 (0.90 L) & 0.00 (0.82 L) & 0.00 (0.93 L) & 0.00 (0.81 L) & 0.00 (0.91 L)\\ \hline
MAP  & 0.00 (0.73 L) & 0.00 (0.85 L) & 0.00 (0.96 L) & 0.00 (0.99 L) & 0.00 (0.85 L) & 0.00 (0.93 L) & 0.00 (1.0 L) & 0.00 (1.0 L)\\ \hline
MRR  & 0.00 (0.70 L) & 0.00 (0.88 L) & 0.00 (0.86 L) & 0.00 (0.98 L) & 0.00 (0.74 L) & 0.00 (0.93 L) & 0.00 (0.98 L) & 0.00 (0.99 L)\\ \hline
& \multicolumn{2}{|c}{\textbf{P5}} & \multicolumn{2}{|c}{\textbf{P6}} & \multicolumn{2}{|c}{\textbf{P7}} & \multicolumn{2}{|c}{\textbf{P8}}\\ \hline
recall@1  & 0.00 (0.74 L) & 0.00 (0.87 L) & 0.00 (0.73 L) & 0.00 (0.50 L) & 0.00 (1.0 L) & 0.00 (1.0 L) & 0.00 (0.70 L) & 0.00 (0.98 L)\\ \hline
recall@5  & 0.00 (0.68 L) & 0.00 (0.77 L) & 0.00 (0.90 L) & 0.00 (0.89 L) & 0.00 (1.0 L) & 0.00 (1.0 L) & 0.00 (0.69 L) & 0.00 (0.97 L)\\ \hline
recall@10  & 0.00 (0.74 L) & 0.00 (0.88 L) & 0.00 (0.71 L) & 0.00 (0.85 L) & 0.00 (0.98 L) & 0.00 (0.99 L) & 0.00 (0.72 L) & 0.00 (0.98 L)\\ \hline
MAP  & 0.00 (0.91 L) & 0.00 (0.93 L) & 0.00 (0.72 L) & 0.00 (0.80 L) & 0.00 (0.96 L) & 0.00 (0.97 L) & 0.00 (0.62 L) & 0.00 (0.74 L)\\ \hline
MRR  & 0.00 (0.87 L) & 0.00 (0.89 L) & 0.00 (0.70 L) & 0.00 (0.84 L) & 0.00 (1.0 L) & 0.00 (0.99 L) & 0.00 (0.70 L) & 0.00 (0.98 L)\\ \hline
& \multicolumn{2}{|c}{\textbf{P9}} & \multicolumn{2}{|c}{\textbf{P10}} & \multicolumn{2}{|c}{\textbf{P11}} & \multicolumn{2}{|c}{\textbf{P12}}\\ \hline
recall@1  & 0.00 (0.81 L) & 0.00 (0.99 L) & 0.00 (0.85 L) & 0.00 (0.99 L) & 0.00 (0.79 L) & 0.00 (0.83 L) & 0.00 (0.84 L) & 0.00 (0.98 L)\\ \hline
recall@5  & 0.00 (0.88 L) & 0.00 (1.0 L) & 0.00 (0.96 L) & 0.00 (0.98 L) & 0.00 (0.76 L) & 0.00 (0.92 L) & 0.00 (0.83 L) & 0.00 (0.96 L)\\ \hline
recall@10  & 0.00 (0.77 L) & 0.00 (0.99 L) & 0.00 (0.60 L) & 0.00 (0.70 L) & 0.00 (0.73 L) & 0.00 (0.94 L) & 0.00 (0.77 L) & 0.00 (0.95 L)\\ \hline
MAP  & 0.00 (0.81 L) & 0.00 (0.99 L) & 0.00 (0.61 L) & 0.00 (0.95 L) & 0.00 (0.79 L) & 0.00 (0.69 L) & 0.00 (0.85 L) & 0.00 (0.98 L)\\ \hline
MRR  & 0.00 (0.76 L) & 0.00 (1.0 L) & 0.00 (0.83 L) & 0.00 (0.99 L) & 0.00 (0.83 L) & 0.00 (0.90 L) & 0.00 (0.78 L) & 0.00 (0.97 L)\\ \hline
\end{tabular}
}
\scriptsize{Note that: The figures X (Y Z) respectively denotes p-value, Cliffs delta, and interpretation of Cliffs delta (i.e., Large (L), Medium (M), Small (S), and Negligible (N))}
\end{table*}

Among the three baselines, the \textit{IR-EM} and \textit{NextBug} employ the \textit{product} field or \textit{component} field to help detect duplicate reports.
We use a different field \textit{test task id} in this experiment.
The reason is that the crowdtesting reports do not have the field \textit{product} or \textit{component},
and \textit{test task id} is the most similar field.
Moreover, we also experiment with four other fields, i.e., \textit{phone type}, \textit{operation system}, \textit{ROM information}, and \textit{network environment} (as shown in Table \ref{tab:report_example}) and the performance is even worse.

The performance of \textit{NextBug} is almost the worst.
This might because the reports from different test tasks
are already distinguishable in their textual descriptions.
Therefore, the utilization of the \textit{test task id} field could not provide extra
information in detecting duplicate reports.
The low performance of \textit{IR-EM} in our crowdtesting reports dataset might due to the similar reason.
As the \textit{test task id} almost could not contribute to duplicate detection, the baseline \textit{IR-EM} degenerates to,
to some extent, the approach of only using textual descriptions (i.e., the TF-IDF and word embedding features, see results in Section \ref{subsec:result_rq2}).

The \textit{DBTM} utilizes term-based and topic-based features for duplicate detection.
The low performance of this baseline might because, unlike the large-scale open source projects, the reports of one crowdtesting project only have very few topics.
The optimal topic number is about 100 to 300 for Eclipse, OpenOffice, and Mozilla \cite{duplicate_topic_model}.
However, the optimal topic number for our experimental projects is about 5 to 10.
% (Note that, the performance in Figure \ref{fig:RQ2} is obtained on the topic number 11, which is a relative optimal value for all projects).
The tiny number of topics cannot effectively help distinguish duplicate reports.

\mybox{Compared with three state-of-the-art and typical baselines, {\tool} significantly and substantially achieves a better performance in terms of all the evaluation metrics for all experimental projects. 
\textit{recall@1} is 0.44 to 0.79, \textit{recall@5} is 0.66 to 0.92, and \textit{MAP} is 0.21 to 0.58 across all experimental projects. 
}

\subsection{Answering RQ2: Necessity}
\label{subsec:result_rq2}

Figure \ref{tab:rq2} presents the \textit{recall@1}, \textit{recall@5}, \textit{recall@10}, \textit{MAP}, and \textit{MRR} for each experimental project for {\tool} and \textit{onlyText}, \textit{onlyImage} (see Section \ref{subsec:experiment_setup} for detail).

We can observe that the performance obtained by \textit{onlyText} or \textit{onlyImage} is worse than {\tool}.
The improvement in \textit{recall@1} of {\tool} is 23\% to 211\% compared with \textit{onlyText}, and 17\% to 319\% compared with \textit{onlyImage}.
The improvement in \textit{MAP} of {\tool} is 31\% to 241\% compared with \textit{onlyText}, and 40\% to 552\% compared with \textit{onlyImage}.

We further conduct Mann-Whitney U Test for the five metrics between {\tool} and \textit{onlyText}, \textit{onlyImage}.
Table \ref{tab:rq2-test} shows the p-value, the Cliff’s delta, and the interpretation of these tests (see Section \ref{subsec:experiment_metric} for details).
We can see that, compared with only using screenshots or textual descriptions (i.e., \textit{onlyText}, \textit{onlyImage}), {\tool} significantly (i.e., p-value for all tests is less than 0.05) and substantially (i.e., Cliff's delta for all tests is large ) achieves a better performance in terms of all the evaluation metrics for all experimental projects. 
This further indicates that only using screenshots or textual descriptions is not effective enough,
and combining these two sources of information is a sensible choice for duplicate crowdtesting report detection.
%This further indicates the necessity of both textual and screenshot in detecting duplicate crowdtesting reports. 

The performance of \textit{onlyImage} is a little lower than the performance of \textit{onlyText}.
This might because the screenshot mainly provides the context-related information. Without the assistance of textual descriptions, the screenshot can not distinguish the duplicate reports in many circumstances.
Moreover, duplicate detection with only textual descriptions can neither achieve equivalent performance with {\tool}, denoting the screenshot plays an indispensable role in detecting duplicate crowdtesting reports.

\mybox{Duplicate detection with only screenshots or textual descriptions achieves significantly and substantially worse performance than {\tool} in all the evaluation metrics for all experimental projects. 
This indicates the necessity of using both screenshots and textual descriptions in detecting duplicate crowdtesting reports. 
}

\subsection{Answering RQ3: Replaceability}
\label{subsec:result_rq3}

\begin{table*}[!t]
\scriptsize
\caption{\textbf{Performance comparison between {\tool} and \textit{noClr}, \textit{noStrc}, \textit{noTF}, \textit{noEmb} (RQ3)}}
\label{tab:rq3}
\centering\scalebox{0.90}{
\vspace{0.05in}
\begin{tabular}{p{0.85cm}|p{0.5cm}|p{0.5cm}|p{0.5cm}|p{0.5cm}|p{0.5cm}|p{1.25cm}|p{0.5cm}|p{0.5cm}|p{0.5cm}|p{0.5cm}|p{0.5cm}|p{1.1cm}|p{0.5cm}|p{0.5cm}|p{0.5cm}|p{0.5cm}|p{0.5cm}|p{1.1cm}}
\hline
& \textbf{SETU} & \textbf{noClr} & \textbf{noStrc} & \textbf{noTF} & \textbf{noEmb} & \textbf{Imprv.} & \textbf{SETU} & \textbf{noClr} & \textbf{noStrc} & \textbf{noTF} & \textbf{noEmb} & \textbf{Imprv.} & \textbf{SETU} & \textbf{noClr} & \textbf{noStrc} & \textbf{noTF} & \textbf{noEmb} & \textbf{Imprv.}  \\ \hline
& \multicolumn{6}{|c}{\textbf{P1}} & \multicolumn{6}{|c}{\textbf{P2}} & \multicolumn{6}{|c}{\textbf{P3}}\\ \hline
recall@1  & \textbf{0.792}  & 0.792 & 0.712 & 0.776 & 0.768 & \textit{0\% - 11\%}  & \textbf{0.649}  & 0.649 & 0.634 & 0.648 & 0.653 & \textit{0\% - 2\%}  & \textbf{0.715}  & 0.657 & 0.663 & 0.626 & 0.663 & \textit{7\% - 14\%} \\ \hline
recall@5  & \textbf{0.872}  & 0.870 & 0.840 & 0.860 & 0.860 & \textit{0\% - 3\%}  & \textbf{0.836}  & 0.820 & 0.817 & 0.830 & 0.807 & \textit{0\% - 3\%}  & \textbf{0.894}  & 0.884 & 0.857 & 0.884 & 0.857 & \textit{1\% - 4\%} \\ \hline
recall@10  & \textbf{0.944}  & 0.928 & 0.928 & 0.940 & 0.928 & \textit{0\% - 1\%}  & \textbf{0.875}  & 0.873 & 0.850 & 0.869 & 0.860 & \textit{0\% - 2\%}  & \textbf{0.915}  & 0.905 & 0.889 & 0.901 & 0.905 & \textit{1\% - 2\%} \\ \hline
MAP  & \textbf{0.280}  & 0.277 & 0.232 & 0.269 & 0.268 & \textit{1\% - 20\%}  & \textbf{0.570}  & 0.573 & 0.537 & 0.563 & 0.555 & \textit{0\% - 6\%}  & \textbf{0.452}  & 0.434 & 0.435 & 0.432 & 0.435 & \textit{3\% - 4\%} \\ \hline
MRR  & \textbf{0.831}  & 0.836 & 0.781 & 0.821 & 0.816 & \textit{0\% - 6\%}  & \textbf{0.736}  & 0.734 & 0.718 & 0.726 & 0.735 & \textit{0\% - 2\%}  & \textbf{0.794}  & 0.755 & 0.747 & 0.737 & 0.757 & \textit{4\% - 7\%} \\ \hline
& \multicolumn{6}{|c}{\textbf{P4}} & \multicolumn{6}{|c}{\textbf{P5}} & \multicolumn{6}{|c}{\textbf{P6}}\\ \hline
recall@1  & \textbf{0.729}  & 0.760 & 0.739 & 0.729 & 0.750 & \textit{-4\% - 0\%}  & \textbf{0.553}  & 0.541 & 0.482 & 0.529 & 0.458 & \textit{2\% - 20\%}  & \textbf{0.722}  & 0.746 & 0.674 & 0.726 & 0.734 & \textit{-3\% - 7\%} \\ \hline
recall@5  & \textbf{0.927}  & 0.906 & 0.875 & 0.916 & 0.864 & \textit{1\% - 7\%}  & \textbf{0.815}  & 0.809 & 0.767 & 0.813 & 0.750 & \textit{0\% - 8\%}  & \textbf{0.871}  & 0.867 & 0.839 & 0.859 & 0.867 & \textit{0\% - 3\%} \\ \hline
recall@10  & \textbf{0.958}  & 0.947 & 0.927 & 0.947 & 0.906 & \textit{1\% - 5\%}  & \textbf{0.922}  & 0.904 & 0.898 & 0.916 & 0.821 & \textit{0\% - 12\%}  & \textbf{0.915}  & 0.903 & 0.891 & 0.915 & 0.895 & \textit{0\% - 2\%} \\ \hline
MAP  & \textbf{0.584}  & 0.607 & 0.559 & 0.583 & 0.566 & \textit{-3\% - 4\%}  & \textbf{0.288}  & 0.287 & 0.247 & 0.281 & 0.253 & \textit{0\% - 16\%}  & \textbf{0.543}  & 0.547 & 0.502 & 0.531 & 0.534 & \textit{0\% - 8\%} \\ \hline
MRR  & \textbf{0.815}  & 0.827 & 0.806 & 0.819 & 0.809 & \textit{-1\% - 1\%}  & \textbf{0.663}  & 0.655 & 0.608 & 0.669 & 0.589 & \textit{0\% - 12\%}  & \textbf{0.792}  & 0.801 & 0.754 & 0.795 & 0.792 & \textit{-1\% - 5\%} \\ \hline
& \multicolumn{6}{|c}{\textbf{P7}} & \multicolumn{6}{|c}{\textbf{P8}} & \multicolumn{6}{|c}{\textbf{P9}}\\ \hline
recall@1  & \textbf{0.542}  & 0.440 & 0.420 & 0.440 & 0.400 & \textit{23\% - 35\%}  & \textbf{0.647}  & 0.612 & 0.603 & 0.636 & 0.578 & \textit{1\% - 11\%}  & \textbf{0.549}  & 0.540 & 0.575 & 0.540 & 0.416 & \textit{-4\% - 31\%} \\ \hline
recall@5  & \textbf{0.666}  & 0.665 & 0.675 & 0.685 & 0.655 & \textit{-2\% - 1\%}  & \textbf{0.849}  & 0.829 & 0.833 & 0.826 & 0.780 & \textit{1\% - 8\%}  & \textbf{0.768}  & 0.746 & 0.781 & 0.742 & 0.648 & \textit{-1\% - 18\%} \\ \hline
recall@10  & \textbf{0.693}  & 0.706 & 0.696 & 0.706 & 0.696 & \textit{-1\% - 0\%}  & \textbf{0.877}  & 0.862 & 0.864 & 0.867 & 0.841 & \textit{1\% - 4\%}  & \textbf{0.811}  & 0.798 & 0.824 & 0.793 & 0.759 & \textit{-1\% - 6\%} \\ \hline
MAP  & \textbf{0.259}  & 0.209 & 0.199 & 0.209 & 0.179 & \textit{23\% - 44\%}  & \textbf{0.321}  & 0.311 & 0.283 & 0.312 & 0.277 & \textit{2\% - 15\%}  & \textbf{0.307}  & 0.304 & 0.315 & 0.295 & 0.236 & \textit{-2\% - 30\%} \\ \hline
MRR  & \textbf{0.565}  & 0.532 & 0.522 & 0.522 & 0.502 & \textit{6\% - 12\%}  & \textbf{0.738}  & 0.718 & 0.703 & 0.726 & 0.672 & \textit{1\% - 9\%}  & \textbf{0.644}  & 0.633 & 0.666 & 0.633 & 0.530 & \textit{-3\% - 21\%} \\ \hline
& \multicolumn{6}{|c}{\textbf{P10}} & \multicolumn{6}{|c}{\textbf{P11}} & \multicolumn{6}{|c}{\textbf{P12}}\\ \hline
recall@1  & \textbf{0.440}  & 0.443 & 0.480 & 0.484 & 0.099 & \textit{-9\% - 344\%}  & \textbf{0.773}  & 0.765 & 0.775 & 0.761 & 0.775 & \textit{0\% - 1\%}  & \textbf{0.719}  & 0.719 & 0.699 & 0.699 & 0.709 & \textit{0\% - 2\%} \\ \hline
recall@5  & \textbf{0.695}  & 0.664 & 0.714 & 0.732 & 0.397 & \textit{-5\% - 75\%}  & \textbf{0.887}  & 0.889 & 0.868 & 0.883 & 0.879 & \textit{0\% - 2\%}  & \textbf{0.927}  & 0.907 & 0.897 & 0.907 & 0.897 & \textit{2\% - 3\%} \\ \hline
recall@10  & \textbf{0.726}  & 0.712 & 0.747 & 0.763 & 0.590 & \textit{-4\% - 23\%}  & \textbf{0.924}  & 0.922 & 0.910 & 0.926 & 0.893 & \textit{0\% - 3\%}  & \textbf{0.948}  & 0.928 & 0.908 & 0.918 & 0.898 & \textit{2\% - 5\%} \\ \hline
MAP  & \textbf{0.219}  & 0.211 & 0.226 & 0.242 & 0.089 & \textit{-9\% - 146\%}  & \textbf{0.450}  & 0.436 & 0.444 & 0.447 & 0.442 & \textit{0\% - 3\%}  & \textbf{0.564}  & 0.554 & 0.544 & 0.544 & 0.524 & \textit{1\% - 7\%} \\ \hline
MRR  & \textbf{0.552}  & 0.554 & 0.573 & 0.588 & 0.249 & \textit{-6\% - 121\%}  & \textbf{0.828}  & 0.823 & 0.812 & 0.821 & 0.824 & \textit{0\% - 1\%}  & \textbf{0.805}  & 0.785 & 0.775 & 0.805 & 0.785 & \textit{0\% - 3\%} \\ \hline

\end{tabular}
}
\end{table*}

\begin{table*}[!t]
\scriptsize
\caption{\textbf{Results of Mann-Whitney U Test between {\tool} and \textit{noClr}, \textit{noStrc}, \textit{noTF}, \textit{noEmb} (RQ3)}}
\label{tab:rq3-test}
\centering\scalebox{0.82}{
\vspace{0.05in}
\begin{tabular}{p{0.9cm}|p{1.3cm}|p{1.35cm}|p{1.35cm}|p{1.35cm}|p{1.3cm}|p{1.35cm}|p{1.3cm}|p{1.35cm}|p{1.3cm}|p{1.35cm}|p{1.35cm}|p{1.3cm}}
\hline
&  \textbf{SETU vs. noClr} & \textbf{SETU vs. noStrc} & \textbf{SETU vs.  noTF} &   \textbf{SETU vs. noEmb} &  \textbf{SETU vs.  noClr} & \textbf{SETU vs.  noStrc} & \textbf{SETU vs.   noTF} &   \textbf{SETU vs.  noEmb} &
\textbf{SETU vs.  noClr} & \textbf{SETU vs.  noStrc} & \textbf{SETU vs.  noTF} &   \textbf{SETU vs. noEmb}\\ \hline
& \multicolumn{4}{|c}{\textbf{P1}} & \multicolumn{4}{|c}{\textbf{P2}} & \multicolumn{4}{|c}{\textbf{P3}}\\ \hline
recall@1  & 0.05 (0.11 N) & 0.00 (0.41 M) & 0.05 (0.11 N) & 0.00 (0.30 S) & 0.71 (-0.0 N) & 0.31 (0.02 N) & 0.48 (0.00 N) & 0.67 (-0.0 N) & 0.00 (0.24 S) & 0.00 (0.29 S) & 0.00 (0.42 M) & 0.00 (0.28 S)\\ \hline
recall@5  & 0.88 (-0.0 N) & 0.00 (0.18 S) & 0.36 (0.02 N) & 0.02 (0.14 N) & 0.02 (0.11 N) & 0.36 (0.01 N) & 0.50 (-0.0 N) & 0.01 (0.12 N) & 0.05 (0.09 N) & 0.00 (0.19 S) & 0.12 (0.06 N) & 0.00 (0.21 S)\\ \hline
recall@10  & 0.37 (0.02 N) & 0.82 (-0.0 N) & 0.71 (-0.0 N) & 0.52 (-0.0 N) & 0.91 (-0.0 N) & 0.04 (0.09 N) & 0.71 (-0.0 N) & 0.28 (0.03 N) & 0.09 (0.07 N) & 0.00 (0.17 S) & 0.00 (0.14 N) & 0.09 (0.07 N)\\ \hline
MAP  & 0.01 (0.15 S) & 0.00 (0.48 L) & 0.02 (0.14 N) & 0.00 (0.27 S) & 0.66 (-0.0 N) & 0.00 (0.29 S) & 0.12 (0.06 N) & 0.00 (0.13 N) & 0.01 (0.12 N) & 0.00 (0.18 S) & 0.06 (0.09 N) & 0.00 (0.15 S)\\ \hline
MRR  & 0.79 (-0.0 N) & 0.00 (0.24 S) & 0.24 (0.05 N) & 0.67 (-0.0 N) & 0.25 (0.03 N) & 0.01 (0.12 N) & 0.24 (0.03 N) & 0.25 (0.03 N) & 0.00 (0.19 S) & 0.00 (0.28 S) & 0.00 (0.26 S) & 0.00 (0.32 S)\\ \hline
& \multicolumn{4}{|c}{\textbf{P4}} & \multicolumn{4}{|c}{\textbf{P5}} & \multicolumn{4}{|c}{\textbf{P6}}\\ \hline
recall@1  & 0.99 (-0.2 N) & 0.89 (-0.1 N) & 0.49 (0.00 N) & 0.99 (-0.3 N) & 0.16 (0.06 N) & 0.00 (0.41 M) & 0.00 (0.20 S) & 0.00 (0.53 L) & 0.61 (-0.0 N) & 0.00 (0.37 M) & 0.31 (0.02 N) & 0.87 (-0.0 N)\\ \hline
recall@5  & 0.16 (0.08 N) & 0.00 (0.23 S) & 0.56 (-0.0 N) & 0.02 (0.16 S) & 0.09 (0.08 N) & 0.00 (0.27 S) & 0.04 (0.10 N) & 0.00 (0.36 M) & 0.06 (0.07 N) & 0.00 (0.18 S) & 0.04 (0.09 N) & 0.96 (-0.0 N)\\ \hline
recall@10  & 0.05 (0.12 N) & 0.00 (0.23 S) & 0.00 (0.20 S) & 0.00 (0.34 M) & 0.09 (0.08 N) & 0.01 (0.14 N) & 0.29 (0.03 N) & 0.00 (0.48 L) & 0.14 (0.05 N) & 0.01 (0.10 N) & 0.84 (-0.0 N) & 0.11 (0.06 N)\\ \hline
MAP  & 0.99 (-0.2 N) & 0.02 (0.17 S) & 0.56 (-0.0 N) & 0.40 (0.02 N) & 0.38 (0.01 N) & 0.00 (0.37 M) & 0.12 (0.07 N) & 0.00 (0.36 M) & 0.53 (-0.0 N) & 0.00 (0.16 S) & 0.39 (0.01 N) & 0.44 (0.00 N)\\ \hline
MRR  & 0.60 (-0.0 N) & 0.41 (0.01 N) & 0.90 (-0.1 N) & 0.18 (0.07 N) & 0.40 (0.01 N) & 0.00 (0.19 S) & 0.99 (-0.1 N) & 0.00 (0.37 M) & 0.55 (-0.0 N) & 0.00 (0.27 S) & 0.52 (-0.0 N) & 0.26 (0.03 N)\\ \hline
& \multicolumn{4}{|c}{\textbf{P7}} & \multicolumn{4}{|c}{\textbf{P8}} & \multicolumn{4}{|c}{\textbf{P9}}\\ \hline
recall@1  & 0.00 (0.57 L) & 0.00 (0.69 L) & 0.00 (0.61 L) & 0.00 (0.83 L) & 0.00 (0.19 S) & 0.00 (0.23 S) & 0.06 (0.06 N) & 0.00 (0.30 S) & 0.01 (0.06 N) & 0.99 (-0.1 N) & 0.00 (0.07 N) & 0.00 (0.72 L)\\ \hline
recall@5  & 0.19 (0.10 N) & 0.57 (-0.0 N) & 0.77 (-0.0 N) & 0.48 (0.00 N) & 0.09 (0.05 N) & 0.09 (0.05 N) & 0.00 (0.13 N) & 0.00 (0.32 S) & 0.00 (0.07 N) & 0.95 (-0.0 N) & 0.00 (0.13 N) & 0.00 (0.56 L)\\ \hline
recall@10  & 0.33 (0.05 N) & 0.55 (-0.0 N) & 0.53 (-0.0 N) & 0.29 (0.06 N) & 0.24 (0.02 N) & 0.06 (0.06 N) & 0.17 (0.03 N) & 0.00 (0.16 S) & 0.00 (0.08 N) & 0.79 (-0.0 N) & 0.00 (0.09 N) & 0.00 (0.32 S)\\ \hline
MAP  & 0.00 (0.53 L) & 0.00 (0.64 L) & 0.00 (0.47 M) & 0.00 (0.69 L) & 0.01 (0.08 N) & 0.00 (0.32 S) & 0.00 (0.11 N) & 0.00 (0.40 M) & 0.00 (0.08 N) & 0.96 (-0.0 N) & 0.00 (0.17 S) & 0.00 (0.63 L)\\ \hline
MRR  & 0.05 (0.18 S) & 0.00 (0.32 S) & 0.00 (0.30 S) & 0.00 (0.34 M) & 0.00 (0.11 N) & 0.00 (0.18 S) & 0.00 (0.11 N) & 0.00 (0.33 M) & 0.03 (0.06 N) & 0.99 (-0.2 N) & 0.42 (0.00 N) & 0.00 (0.64 L)\\ \hline
& \multicolumn{4}{|c}{\textbf{P10}} & \multicolumn{4}{|c}{\textbf{P11}} & \multicolumn{4}{|c}{\textbf{P12}}\\ \hline
recall@1  & 0.78 (-0.0 N) & 0.99 (-0.3 N) & 0.99 (-0.3 N) & 0.00 (1.0 L) & 0.09 (0.08 N) & 0.79 (-0.0 N) & 0.17 (0.06 N) & 0.66 (-0.0 N) & 0.13 (0.05 N) & 0.00 (0.13 N) & 0.00 (0.12 N) & 0.02 (0.10 N)\\ \hline
recall@5  & 0.00 (0.19 S) & 0.99 (-0.1 N) & 0.99 (-0.2 N) & 0.00 (0.98 L) & 0.45 (0.00 N) & 0.10 (0.07 N) & 0.20 (0.05 N) & 0.15 (0.06 N) & 0.00 (0.16 S) & 0.00 (0.20 S) & 0.00 (0.12 N) & 0.00 (0.21 S)\\ \hline
recall@10  & 0.11 (0.06 N) & 0.98 (-0.1 N) & 0.99 (-0.1 N) & 0.00 (0.57 L) & 0.53 (-0.0 N) & 0.57 (-0.0 N) & 0.80 (-0.0 N) & 0.06 (0.09 N) & 0.02 (0.10 N) & 0.00 (0.20 S) & 0.00 (0.20 S) & 0.00 (0.22 S)\\ \hline
MAP  & 0.15 (0.05 N) & 0.96 (-0.0 N) & 0.98 (-0.1 N) & 0.00 (0.90 L) & 0.39 (0.01 N) & 0.58 (-0.0 N) & 0.31 (0.03 N) & 0.56 (-0.0 N) & 0.27 (0.03 N) & 0.00 (0.15 S) & 0.00 (0.13 N) & 0.00 (0.25 S)\\ \hline
MRR  & 0.53 (-0.0 N) & 0.97 (-0.1 N) & 0.99 (-0.2 N) & 0.00 (0.99 L) & 0.97 (-0.1 N) & 0.17 (0.06 N) & 0.29 (0.03 N) & 0.37 (0.01 N) & 0.00 (0.13 N) & 0.00 (0.16 S) & 0.49 (0.00 N) & 0.00 (0.13 N)\\ \hline
\end{tabular}
}
\scriptsize{Note that: The figures X (Y Z) respectively denotes p-value, Cliffs delta, and interpretation of Cliffs delta (i.e., Large (L), Medium (M), Small (S), and Negligible (N))}
\end{table*}

Figure \ref{tab:rq3} presents the \textit{recall@1}, \textit{recall@5}, \textit{recall@10}, \textit{MAP}, and \textit{MRR} for each experimental project for {\tool} and \textit{noClr}, \textit{noStrc}, \textit{noTF}, \textit{noEmb} (see Section \ref{subsec:experiment_setup} for detail).

We can observe that, in most circumstances, performance obtained by {\tool} is better than or equal with the performance obtained by using three features (i.e., \textit{noClr}, \textit{noStrc}, \textit{noTF}, or \textit{noEmb}).
In rare cases, the performance obtained by using three features is a little better than the performance of {\tool}. 

We further conduct Mann-Whitney U Test for the five metrics between {\tool} and \textit{noClr}, \textit{noStrc}, \textit{noTF}, \textit{noEmb}.
Table \ref{tab:rq3-test} shows the p-value, the Cliff’s delta, and the interpretation of these tests (see Section \ref{subsec:experiment_metric} for details).
There are totally 240 tests (12 projects, 4 types of features, 5 evaluation metrics).
Among them, in 35\% (83/240) tests, the performance obtained by {\tool} is significantly (i.e., p-value is less than 0.05) and substantially (i.e., Cliff’s delta is not negligible, i.e., small in 21\% tests, median in 5\% tests, or large in 9\% tests) better than the performance of using three features. 
In other 65\% (157/240) tests, the performance obtained by {\tool} demonstrates negligible difference (i.e., Cliff's delta is negligible although some p-value is less than 0.05) with the performance of using three features. 
In none of the 240 tests, the performance of using three features is significantly and substantially better than the performance of {\tool}.
This further indicates our proposed approach of combining these four types of features is effective.

Among the four experiments with three types of features, we can see that \textit{noEmb} achieves relatively worse results. 
This might because the word embedding feature focuses on the relationship of terms by considering the context they appear. 
Without this feature (i.e., \textit{noEmb}), simply matching the occurrence of terms, which is done by TF-IDF feature, cannot effectively detecting the duplicate reports. 
This implies word embedding feature is the least replaceable feature, i.e., the performance would undergo a relatively large decline if removing this feature. 

We can also see that \textit{noClr} achieves relatively better results among the four experiments with three types of features. 
This indicates image color feature is the most replaceable feature, i.e., the performance would undergo a relatively small decrease if removing this feature. 
This also implies that image color feature would sometimes bring noise in the duplicate detection.
We will present further discussion in Section \ref{subsec:dis_image}.

\mybox{Word embedding is the least replaceable feature, while image color is the most replaceable feature. In addition, duplicate detection with all the four features can achieve relatively best performance. }

\section{Discussion}
\label{sec:discussion}

\subsection{Further Exploration of Image Features}
\label{subsec:dis_image}

We have mentioned that \textit{noClr} (i.e., duplicate detection without image color feature) can sometimes achieve a slightly better performance than {\tool}, e.g., in P4, P6 (see details in Table \ref{tab:rq3}).
This indicates that image color feature would bring noise to the duplicate detection in some cases.

We have examined the screenshots of the experimental projects, and found that, in these projects (i.e., P4, P6), screenshots of different functionalities (i.e., denoting non-duplicate reports) can sometimes have very similar color distribution (as Figure \ref{fig:color-2} shows).
This is somehow coincident with our common sense, because many apps pursue a unified interface design to make it more user-friendly. 
Under this case, the non-duplicate reports might share very similar image color features, and removing this feature (i.e., \textit{noClr}) can instead increase the detection performance. 

More than that, we also noticed that, in some projects as P9, P10, the performance achieved by \textit{noStrc} (i.e., duplicate detection without image structure feature) is a little better than {\tool}.

We further examined the screenshots of these two projects (i.e., P9, P10), and found that, the images' structure for different functionalities (i.e., denoting non-duplicate reports) 
can sometimes exert no much difference, just like Figure \ref{fig:structure-2} shows. 
In this case, the non-duplicate reports would share very similar image structure features, and removing the structure feature (i.e., \textit{noStrc}) can instead increase the detection performance. 

The above analysis indicates that image color feature can occasionally bring noise and removing it can increase the performance for duplicate detection, and so does the image structure feature. 
We have also analyzed the reasons and found that if non-duplicate reports tend to share a similar color distribution, removing image color feature can achieve a better performance; this also holds true for image structure feature when non-duplicate reports usually share a similar image structure. 

This motivates us to conduct feature selection for specific crowdtesting projects to further improve the duplicate detection performance. 
In practice, this can be done based on the received reports of the project during the crowdtesting process.
One can examine whether a large number of non-duplicate reports share the similar color distribution or structure, so as to suggest which features to be used for detecting the new-coming reports for this specific project. 

\subsection{Advantage of Hierarchical Approach}
\label{subsec:dis_image}

Our proposed {\tool} employs a hierarchical algorithm to detect duplicate reports, i.e., first classify the reports into two classes, and rank them separately.
Under {\tool}, the class information indicates different levels of certainty of being duplicates. 
In detail, the first class contains the reports whose screenshot similarity is large enough, thus they can be more likely to be the duplicates of the query report. 
On the contrary, the second class contains the reports whose screenshots are quite different with the query report, thus they are less likely to be the duplicates of the query report. 

Therefore, when we provide the users with the duplicate detection results, these class information can give more insights of the detection.
For example, we encourage the users put more focus on the first class since they have large probability to be the true duplicates. 
For the second class, the users only need to glance off the reports to examine whether they are the true duplicates, especially when there have been a large number of recommended reports in the first class.
We believe our two-classes hierarchical approach can provide more practical assistance for duplicate detection in real-world crowdtesting practice. 
Meanwhile, existing approaches treated all the reports as a whole, thus could not provide this kinds of guidelines.

\subsection{Alternative Combination Manner}
\label{subsec:dis_alternative}

The motivating examples in Section \ref{subsec:background_motivation} indicate that the role of screenshots is mainly to demonstrate the context-related information, while the role of textual descriptions is to provide detailed illustration of the reported problem.  
This is why our proposed approach {\tool} first uses screenshot similarity to filter the reports to first class and conducts the ranking separately. 
However, one may still argue that other combination manners could achieve better performance.  
We did conduct experiments to explore whether other combination manners could outperform {\tool} in duplicate detection. 
Due to word limit, we provide the detailed results and analysis, as well as the experimental setup in the external link\footnote{https://github.com/wangjunjieISCAS/ISTDuplicateDetection}.
Here, we briefly summarize the main findings. 

We had designed three new combination manners, i.e., \textit{addCmb}, \textit{multiplyCmb}, and \textit{textFirst}. 
Specifically, \textit{addCmb} denotes adding screenshot similarity and textual similarity as one similarity value and ranking the reports based on it, which is a straight-forward manner. 
\textit{multiplyCmb} denotes multiplying the screenshot similarity with textual similarity as one similarity value and ranking the reports based on it, which is borrowed from \cite{duplicate_embedding}.
\textit{textFirst} denotes treating the reports with high textual similarity as the first class and ranking them with the screenshot similarity (the second class is treated as {\tool} does).

We further conduct Mann-Whitney U Test for the five metrics between {\tool} and \textit{addCmb}, \textit{multiplyCmb}, \textit{textFirst}.
There are totally 180 tests (12 projects, 3 alternative combination manners, 5 evaluation metrics).
Among them, in 73\% (131/180) tests, the performance obtained by {\tool} is significantly (i.e., p-value is less than 0.05) and substantially (i.e., Cliff’s delta is not negligible, i.e., small in 30\% tests, median in 15\% tests, or large in 28\% tests) better than the performance of other combination manners. 
In other 27\% (49/180) tests, the performance obtained by {\tool} demonstrates negligible difference (i.e., Cliff's delta is negligible although some p-value is less than 0.05) with the performance of other combination manners. 
In none of the 180 tests, the performance of alternative combination manners is significantly and substantially better than the performance of {\tool}.
To summarize, the combination manner proposed in {\tool} can achieve relatively highest performance than other alternative combination manners.

\subsection{Threats to Validity}
\label{subsec:dis_threats}

The external threats concern the generality of this study.
First, our experiment data consists of {\projects} projects collected from one of the Chinese largest crowdtesting platforms.
We can not ensure that the results of our study could generalize beyond this environment in which it was conducted.
However, the various domains of projects and size of data relatively reduce this risk.
Second, all crowdtesting reports investigated in this study are written in Chinese, and we cannot assure that similar results can be observed on crowdtesting projects in other languages.
But this is alleviated due to the fact that we did not conduct semantic comprehension, but rather simply tokenize sentence and use word as token for experiment.

Internal validity of this study mainly questions the selection and implementation of baselines.
Because there is no approach for duplicate \textit{crowdtesting report} detection, we can only employ the approaches for duplicate \textit{bug report} detection.
Moreover, as the crowdtesting reports do not have the \textit{product} and \textit{component} fields as the original approaches, we employ the most similar field (i.e., \textit{test task id}) for substitution.
In addition, as these three baseline approaches are not publicly available, we implement our own versions rigorously following the steps described in their papers.

Construct validity of this study mainly questions the data processing method.
We rely on the stored duplicate labels of crowdtesting reports to construct the ground truth.
However, this is addressed to some extent due to the fact that testers in the company have no knowledge that this study will be performed for them to artificially modify their labeling.
Besides, we have verified its validity through random sampling and relabeling.

\section{Conclusion}
\label{sec:conclusion}

In this work, we propose {\tool}, which combines the information from both the screenshots and the textual descriptions to detect duplicate crowdtesting reports.
We evaluate the effectiveness of {\tool} on {\projects} commercial projects with {\reports} reports from one of the Chinese largest crowdtesting platforms, and the results are promising.

Note that this paper is just the starting point of the work in progress.
We are closely collaborating with {\company} crowdtesting platform and planning to deploy
the proposed duplicate crowdtesting report detection approach online.
The feedback from real-world case studies will further validate the effectiveness,
as well as guide us in improving {\tool}.
For the future work,
we would like to explore other features, as well as conduct intelligent selection of optimal features to further
improve the duplicate detection performance.

\section{Acknowledgments}
This work is supported by the National Natural Science Foundation of China under grant No.61602450, No.6143200, and China Scholarship Council.
We would like to thank the
testers in Baidu for their great efforts in supporting this work.


\section{Reference}

\bibliographystyle{reference}

% \bibliography{sigproc} % don’t import bib

% \bibliographystyle{reference}
% \bibliography{reference.bib}






% \end{thebibliography}

\end{document}